\documentclass[showpacs,twocolumn,aps,floatfix,superscriptaddress,noshowpacs]{revtex4}
\usepackage[mathlines]{lineno}

\usepackage{dsfont}
\usepackage{amsmath,amssymb,graphicx,bm,color,mathrsfs,verbatim,epstopdf,dcolumn,cancel}

\graphicspath{{figs/},{paper\_figs/}}

\usepackage{hyperref}
\hypersetup{ 
	colorlinks   =  true
}

\begin{document}

\title{Asymptotic Prethermalization in Periodically Driven Classical Spin Chains}

\author{Owen Howell}
\email{olh20@bu.edu}
\affiliation{Department of Physics, Boston University, 590 Commonwealth Ave., Boston, MA 02215, USA}

\author{Phillip Weinberg}
\affiliation{Department of Physics, Boston University, 590 Commonwealth Ave., Boston, MA 02215, USA}

\author{Dries Sels}
\affiliation{Department of Physics, Boston University, 590 Commonwealth Ave., Boston, MA 02215, USA}
\affiliation{Department of Physics, Harvard University, 17 Oxford st., Cambridge, MA 02138, USA}
\affiliation{Theory of quantum and complex systems, Universiteit Antwerpen, B-2610 Antwerpen, Belgium}

\author{Anatoli Polkovnikov}
\affiliation{Department of Physics, Boston University, 590 Commonwealth Ave., Boston, MA 02215, USA}

\author{Marin Bukov}
\email{mgbukov@berkeley.edu}
\affiliation{Department of Physics, University of California, Berkeley, CA 94720, USA}

\begin{abstract}
We reveal a continuous dynamical heating transition between a prethermal and an infinite- temperature stage in a clean, chaotic periodically driven classical spin chain. The transition time is a steep exponential function of the drive frequency, showing that the exponentially long-lived prethermal plateau, originally observed in quantum Floquet systems, survives the classical limit. Even though there is no straightforward generalization of Floquet's theorem to nonlinear systems, we present strong evidence that the prethermal physics is well described by the inverse-frequency expansion. We relate the stability and robustness of the prethermal plateau to drive-induced syn- chronization not captured by the expansion. Our results set the pathway to transfer the ideas of Floquet engineering to classical many-body systems, and are directly relevant for photonic crystals and cold atom experiments in the superfluid regime.
\end{abstract}
	
\date{\today}
	
	
\maketitle
	
\graphicspath{
	{allpaperfigs/},
}

Periodically-driven systems are currently experiencing an unprecedented revival of interest through theoretical and experimental design of novel states of matter. Commonly known as \emph{Floquet engineering}, this approach has enjoyed success in the regime of high driving frequency, where it has been appreciated as a useful tool to ascribe novel properties to otherwise trivial static Hamiltonians~\cite{goldman2014periodically,eckardt2017atomic,bukov2015universal}. Prominent examples include the Kapitza pendulum~\cite{Kapitza}, cold-atom realisations of topological~\cite{oka_09,kitagawa_11,jotzu_14,aidelsburger_14,flaeschner_15,tarnowski2017characterizing,tarnowski_17,aidelsburger2017artificial} and spin-dependent~\cite{jotzu_15} bands, artificial gauge fields~\cite{struck_11,struck_12,hauke_12,struck_13,aidelsburger_13,miyake_13,atala_14,kennedy_15,bukov_SWT}, spin-orbit coupling~\cite{galitski_13,jimenez-garcia_15}, enhanced magnetic correlations~\cite{gorg2017enhancement}, synthetic dimensions~\cite{celi_13,stuhl2015visualizing,mancini2015observation}, and photonic topological insulators~\cite{rechtsman_13,hafezi_14,mittal_14}.

The applicability of Floquet engineering requires the ability to prepare the periodically driven system in the corresponding Floquet state~\cite{desbuquois_17,weinberg2017adiabatic,novicenko_17,bukov2018reinforcement}, and the stability of the system to detrimental heating~\cite{bilitewski_15,bilitewski_15b,mweinberg_15,reitter_17,nager2018parametric,boulier2018parametric}. Presenting a major bottleneck at the forefront of present-date experimental research, heating processes play an important role in many-body Floquet systems, and understanding the underlying physics is expected to offer significant advances in the field. Unlike single-particle quantum systems, such as the kicked rotor~\cite{fishman_82} and weakly-interacting bosonic models~\cite{lellouch_17,lellouch_18}, it is believed that generic isolated clean periodically-driven quantum many-body systems heat up to an infinite-temperature state~\cite{dalessio_14,lazarides_14,bar2017absence,moessner2017equilibration,citro2015dynamical,seetharam_18}, although the debate is not fully settled~\cite{prosen_98a,prosen_98b,prosen_99,prosen_02,dalessio2013many,haldar2018onset}. Heating rates have been shown to be at least exponentially suppressed in the drive frequency~\cite{abanin_15,mori_15}.

\begin{figure}[t!]
	\centering
	\includegraphics[width=1.0\columnwidth]{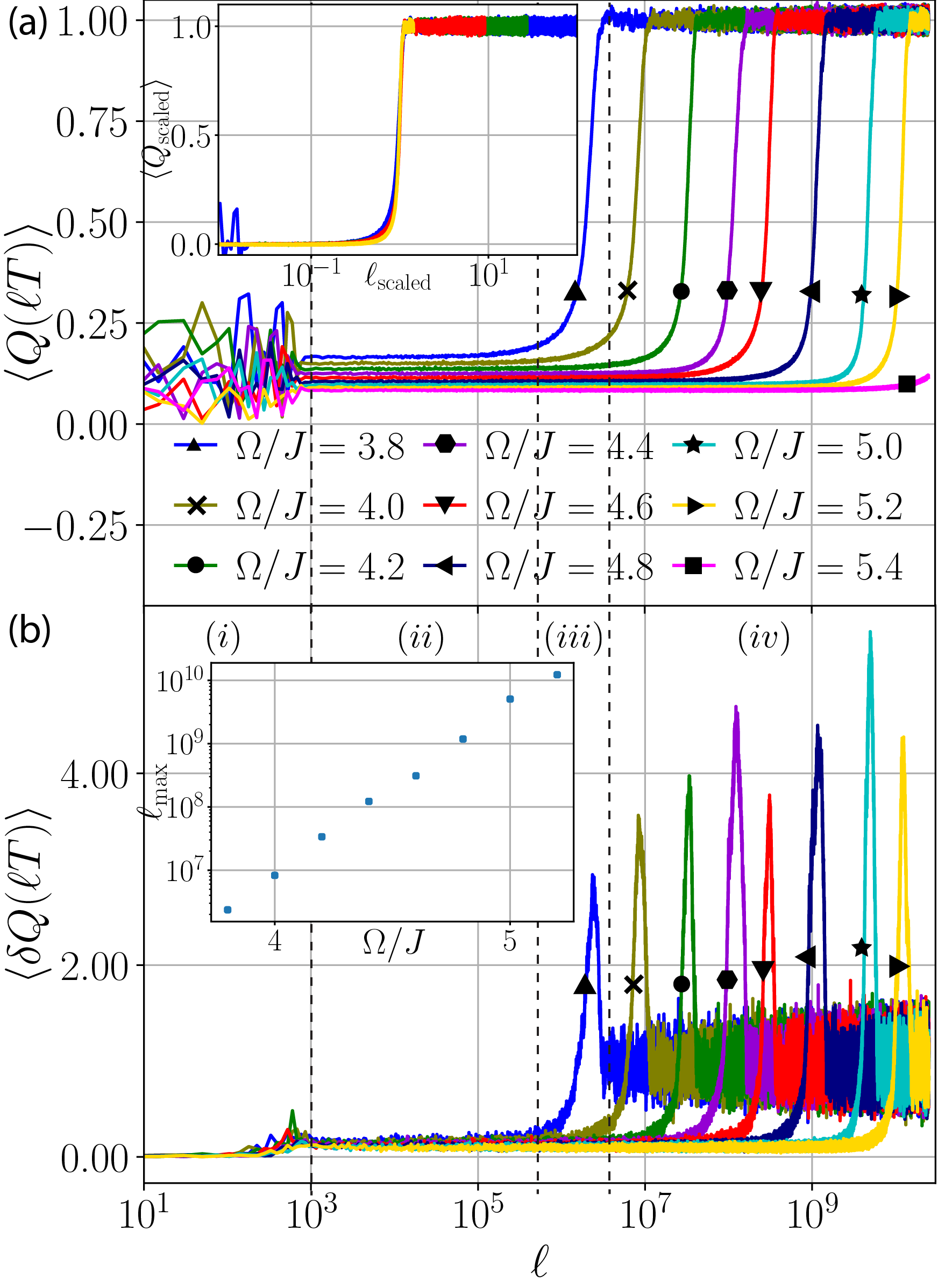}
	\caption{Noise-averaged energy (a) and energy variance (b) as a function of the number of driving cycles $\ell$. Insets: rescaled energy curves (a) by the position of the peak in the variance curves (b), reveal a dynamical heating transition from a prethermal stage to an infinite-temperature stage in the limit $\Omega,\ell \to \infty$. The dashed vertical lines mark the four stages of evolution for $\Omega/J=3.8$ [see text]. See Fig.~\ref{fig:stgd_mag} for parameters.}
	\label{fig:heating_transition}
\end{figure}

In this paper, we present a numerical study of thermalisation in a clean, globally-driven, isolated \emph{classical} spin chain, reaching times beyond the astronomical $10^{10}$ driving cycles. We find that, the dynamics falls into four stages, see Fig.~\ref{fig:heating_transition}: an initial transient $(i)$ during which the system exhibits \emph{constrained} thermalization to a finite energy density set by the initial ensemble, $(ii)$ a frequency-controlled long-lived prethermal plateau~\cite{bukov_15_prl,abanin_15,weidinger2017floquet,peronaci2017resonant} ideally suited for Floquet engineering, followed by $(iii)$ a late crossover governed by \emph{unconstrained} thermalisation to $(iv)$ a featureless infinite-temperature state. Our analysis reveals the existence of a dynamical heating transition between a prethermal plateau and an infinite-temperature stage at infinite times and infinite drive frequencies. The prethermal plateau is described by the inverse-frequency expansion originally developed for quantum systems, with improving agreement at increasing first few orders of the expansion~\cite{supplementary}. This allows to transfer the machinery of Floquet engineering directly to classical many-body models despite the absence of a Floquet theorem for systems governed by nonlinear equations of motion. Focusing on the prethermal plateau, we demonstrate that a key role for its exponentially long duration is played by drive-induced synchronization. In particular, stopping the periodic drive and then restarting it de-synchronizes the system and strongly increases the heating rate.

We report that the time scale for a classical system to leave a small corner of phase space around the initial state and heat up to an infinite-temperature state, scales exponentially with the driving frequency [Fig.~\ref{fig:heating_transition}]. A related problem on ergodicity in classical systems was addressed in a pioneering numerical study by Fermi, Pasta, Ulam and Tsingou, and gave first evidence for a parametrically slow thermalisation in a system of coupled classical oscillators~\cite{fermi_55, Danieli_16}, puzzling the community for the past few decades. For systems with finite number of degrees of freedom, theorems in Hamiltonian mechanics have been proven showing that the motion of action variables in nearly-integrable systems remains confined to a small region of phase space until exponentially long times, controlled by the integrability breaking parameter~\cite{nekhoroshev1971behavior,moser_55,littlewood1959equilateral,poschel1993nekhoroshev}. This behaviour is accompanied by sub-diffusion, as reported for a system of periodically-kicked coupled pendula~\cite{pettini_90,konishi1990diffusion,notarnicola2017localization,rajak_18}. Chaotic many-body dynamics in periodically-kicked spin chains has been studied using a classical Loschmidt echo approach~\cite{veble_04,veble_05}.


\emph{Model.---}Consider a classical Ising chain with periodic boundary conditions, described by the energy function
\begin{equation} \label{eq:1}
H(t) = 
\begin{cases}
\sum_{j=1}^{N} J S_{j}^{z} S_{j + 1}^{z} + h S_{j}^{z} & \text{for}\ t\in [0,T/2]\ \mathrm{mod}\ T  \\
\sum_{j=1}^{N} g S_{j}^{x}  & \text{for}\ t\in [T/2,T]\ \mathrm{mod}\ T 
\end{cases}
\nonumber
\end{equation}
where $J$ denotes the nearest-neighbour interaction strength, while $h$ and $g$ are the magnetic field strengths along the $z$ and $x$-directions, respectively. The spin [or rotor] variable $\vec{S}_j$, $\vert\vec{S}_j\vert=1$, on site $j$ satisfies the Poisson bracket relation $\{S^\mu_i,S^\nu_j\}=\delta_{ij}\varepsilon^{\mu\nu\rho}S^\rho_j$, with $\varepsilon^{\mu\nu\rho}$ the fully antisymmetric tensor.

The time dependence arises due to periodic switching of two time-independent Hamilton functions, for a duration of $T/2$ each, with frequency $\Omega=2\pi/T$. The time evolution of the system is governed by Hamilton's EOM $\dot{S}^\mu_j(t) =\{S^\mu_j,H(t)\}$. Interested in the long-time thermalisation properties, we focus on stroboscopic evolution. Integrating the EOM over one total period $T$, the evolved state is obtained from a successive application of a discrete map $\vec S_j(\ell T) = \left[\tau_2\circ\tau_1\right]^\ell(\vec S_j(0))$,
with $\ell\in\mathbb{N}$ counting the driving cycles. During the first half-period, the time evolution follows the \emph{non-linear} rotation $\tau_1$ about the $z$-axis:
\begin{equation}
\tau_1(\vec S_j) =
\begin{bmatrix}
S_{j}^{x}\cos(\kappa_{j}T/2)-S_{j}^{y}\sin(\kappa_{j}T/2)\\S_{j}^{x}\sin(\kappa_{j}T/2)+S_{j}^{y}\cos(\kappa_{j}T/2)\\S_{j}^{z}
\end{bmatrix}
\end{equation}
with spin-dependent natural frequency of rotation $\kappa_{j} = J(S_{j-1}^{z}+S_{j+1}^{z}) +h$. The dynamics in the second half-period follows the rotation $\tau_2$ about the $x$-axis:
\begin{equation}
\tau_2(\vec S_j) =
\begin{bmatrix}
S_{j}^{x}\\S_{j}^{y}\cos(gT/2)-S_{j}^{y}\sin(gT/2)\\S_{j}^{y}\sin(gT/2) + S_{j}^{z}\cos(gT/2)
\end{bmatrix}
\end{equation}
The map $\tau_2\circ\tau_1$ is the nonlinear classical analogue of the quantum Floquet unitary.

Motivated by experiments which study Floquet-engineered ordered states at high drive frequencies, we prepare the system at time $t=0$ in the lowest-energy state (i.e.~the ground state, GS) of the time-averaged Hamiltonian
\begin{equation}
\label{eq:Have}
H_\mathrm{ave} \equiv  H_F^{(0)} = \frac{1}{2}\sum_{j=1}^{N} J S_{j}^{z} S_{j+1}^{z} + h S_{j}^{z} + g S_{j}^{x},
\end{equation}
with energy density $E_\mathrm{GS}(h/J,g/J)/N\!\approx\!-1.235J$ for $g/J\!=\!0.9045$, $h/J\!=\!0.809$. Whenever $J,h,g$ have equal sign and are of the same order of magnitude, the GS features antiferromagnetic (AFM) order w.r.t.~a direction in the $xz$-plane, parametrized by the azimuthal angle $\theta$. Making use of translational invariance, one can determine the value of $\theta$ which minimizes the energy $E_\mathrm{GS}(\theta)$. Translational invariance constrains the GS evolution to be uniquely described by two coupled spin degrees of freedom, corresponding to the AFM unit cell. To bring out the many-body character of the model, we add small noise to the azimuthal angle of each spin, independently drawn from a uniform distribution over $[-\pi/100,\pi/100]$, which breaks translational symmetry and allows for thermalisation. The quantities we consider are averaged over an ensemble of $100$ noisy initial state realizations. We verified that the long-time dynamics is independent of the strength of the noise~\cite{supplementary}, provided the latter remains small enough to not significantly change the energy of the initial state. In the following, we denote by $\langle\cdot\rangle$ the average over the ensemble of noise realizations.


\emph{Heating Transition.---}Compared to classical systems, studies on thermalising dynamics in quantum models feature serious deficiencies, due to significant finite-size effects inherent to state-of-the-art exact diagonalization simulations. Since energy absorption is known to happen through Floquet many-body resonances~\cite{bukov_15_res,claeys2017breaking,claeys2017spin}, (i), their density depends strongly on the drive frequency at any fixed many-body bandwidth. As the bandwidth scales linearly with the system size $N$, this puts an upper bound on $\Omega$ for the system to be in the many-body regime. This also limits the occurrence of higher-order absorption processes, reducing the overall capacity for energy absorption. (ii) Low-energy initial states, whose energy level spacing does not follow the $2^{-N}$ scaling of the bulk, further restrict the appearance of resonances. However, these issues are intrinsic to quantum models and none of them is problematic in periodically-driven classical systems. The classical energy manifold is continuous allowing for excitations at all energies, and one can easily reach system sizes of several hundred spins, which mitigates the constraint on the reliable upper bound for the driving frequency by a few orders of magnitude. Nevertheless, studying classical systems comes at a notable price: one cannot access the infinite-time behavior, and is thus limited to finite times.

Often, experiments in Floquet engineering are designed to study the GS of the infinite-frequency Hamiltonian~\eqref{eq:Have}. Therefore, $H_\mathrm{ave}$ constitutes a natural observable to measure the excess energy pumped into the system from the drive. Let us define the dimensionless expected energy and energy variance~\cite{dalessio2013many,bukov_15_res}, over the initial ensemble of noisy AFM states:
\begin{eqnarray}
\langle Q(\ell T) \rangle \!\!\!&\equiv&\!\!\! \langle Q^{(0)}(\ell T)\rangle \!=\! \frac{ \langle H_{\mathrm{ave}}[\{\vec S_j(\ell T)\}] \rangle - E_\mathrm{GS} }{\langle H_{\mathrm{ave}}\rangle_{\beta = 0} - E_\mathrm{GS}}\in [0,1], \nonumber \\
\langle \delta Q(\ell T) \rangle \!\!\!&=&\!\!\!\sqrt{ \frac{ \langle H^{2}_\mathrm{ave}[\{\vec S_j(\ell T)\}]\rangle - \langle H_\mathrm{ave}[\{\vec S_j(\ell T)\}]\rangle^{2}} {\langle H_{\mathrm{ave}}^2\rangle_{\beta = 0} - \langle H_{\mathrm{ave}}\rangle_{\beta = 0}^{2}} }. 
\label{eq:Q_ave}
\end{eqnarray}
The normalization is chosen w.r.t.~an infinite-temperature ensemble, where each spin points at a random direction, and hence $\langle H_{\mathrm{ave}}\rangle_{\beta\!=\! 0}\!=\! 0$ and $\langle H_{\mathrm{ave}}^2\rangle_{\beta \!= \! 0}\!=\! N/3  (J^{2}/3 \!+\! h^{2} \!+\! g^{2})$. Initializing the system in the ensemble of noisy AFM states, we have $\langle Q(\ell T) \rangle \approx 0$ if the system does not absorb energy, and $\langle Q(\ell T) \rangle \!=\! 1 $ whenever the ensemble is heated to infinite temperature.

There are four stages in the evolution of the system, see Fig.~{\ref{fig:heating_transition}. Notice that the time [cf.~stage $(iii)$] between the pre-thermal plateau and the infinite-temperature state, corresponding to maximum energy variance: $\ell_\mathrm{max}(\Omega) = \mathrm{argmax}_{\ell} \langle \delta Q(\ell T)\rangle$, scales \emph{exponentially}~\cite{machado2017exponentially} with the driving frequency $\Omega$, c.f.~Fig.~{\ref{fig:heating_transition}b (inset). Our numerical study indicates that for $h/J \!=\! 0.0$ the heating time can be parametrized as $\ell_\mathrm{max}(\Omega)\!=\!r(g/J) \exp[-\gamma(g/J)\times\Omega/J]$, with $\gamma(g/J)$ a slowly-varying function of $g/J$, and $r(g/J)\propto (g/J)^{\alpha}$, $\alpha \!\approx\! -2.12$, within the entire range of existence of the prethermal stage~\cite{supplementary}. Thus, we can rescale the $\langle Q(\ell T) \rangle$ curves with respect to the energy in the beginning of the prethermal plateau $(ii)$:
\begin{equation}
\langle Q_\mathrm{scaled}(\ell_\mathrm{scaled}T) \rangle =  \frac{ \langle Q( \ell_\mathrm{scaled}T ) \rangle - \langle Q\rangle_\mathrm{prethermal} }{ \langle Q \rangle_{\beta = 0} - \langle Q\rangle_\mathrm{prethermal}},
\end{equation}
where $\ell_\mathrm{scaled} =  \ell/\ell_\mathrm{max}(\Omega)$. Figure~\ref{fig:heating_transition}a (inset) shows the collapse in the energy absorption curves with increasing frequency, whereas the width of the peak in $\delta Q(\ell_\mathrm{scaled})$ stays constant~\cite{supplementary}. This behavior has remarkable similarities with the continuous phase transition observed in the 1d Ising model at zero temperature, where the squared magnetization plays the role of energy absorption, while inverse temperature and system size are analogous to drive frequency and time, respectively~\cite{supplementary}; the correlation length corresponds to the heating time. This analogy suggests that heating may happen through a continuous phase transition in the limit $\Omega,t\to\infty$. At moderate frequencies and times, relevant for experiments, we find a sharp crossover instead.

\begin{figure}[t!]
	\centering
	\includegraphics[width=1.0\columnwidth]{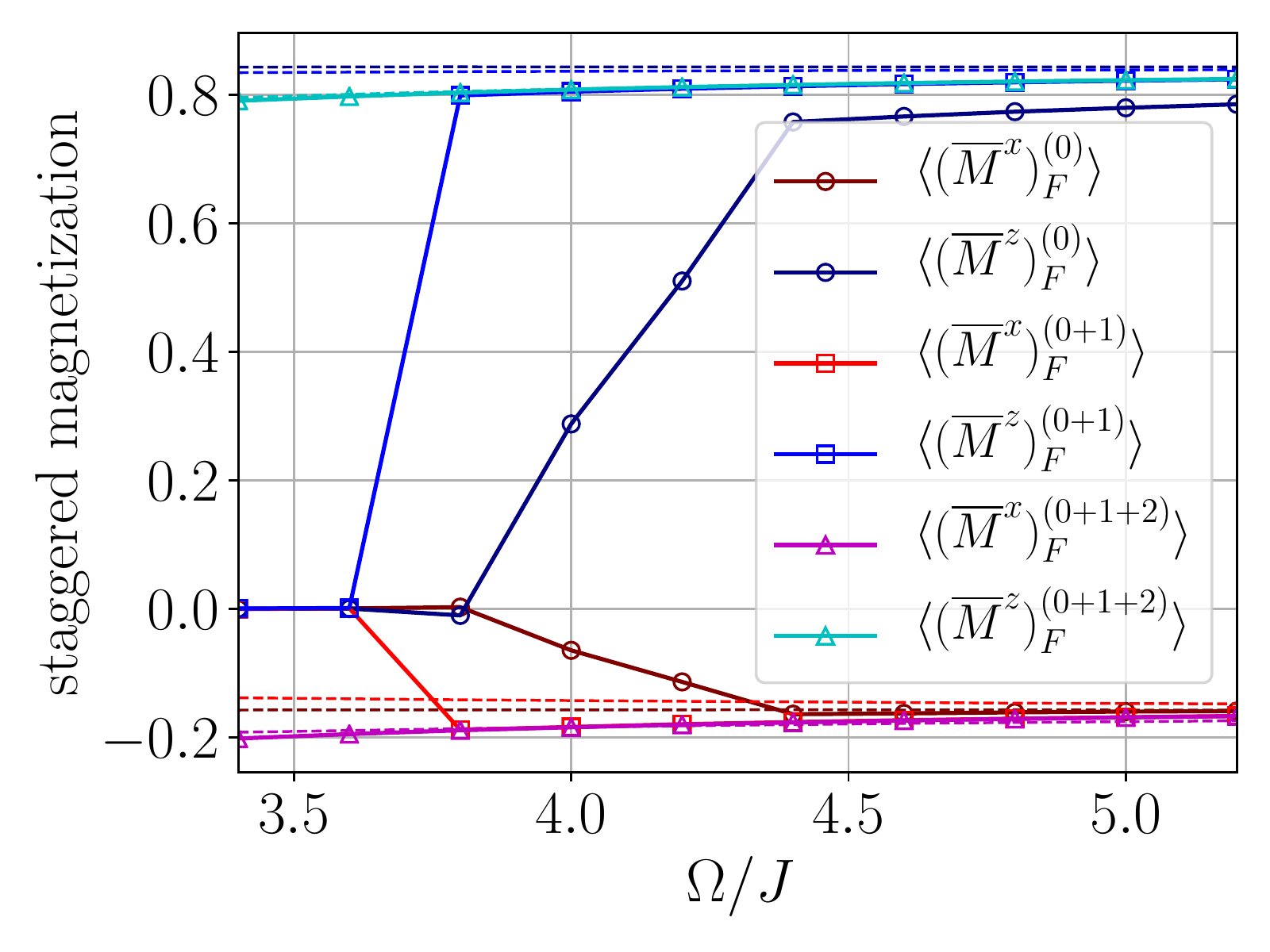}
	\caption{Staggered magnetization $ \langle (  \overline{M}^{\alpha})_{F}^{(0+1+..+m)} \rangle$ in the prethermal plateau to different order $m$ in the ME (solid lines), compared to its value in the initial GS of the corresponding Floquet Hamiltonian (dashed). The parameters are $g/J=0.9045$, $h/J=0.809$, $N_T=10^{6}$, and $N=100$. Every point is averaged over an ensemble of $100$ noise realisations.}
	\label{fig:stgd_mag}
\end{figure}
		
\emph{Prethermal Regime.---}The prethermal plateau plays a crucial role in strongly-interacting Floquet systems because it offers a stable window to experimentally realize novel many-body states of matter. We demonstrate that the inverse-frequency expansion can be used to gain a better understanding of the prethermal plateau, and present compelling evidence that this stage of evolution is captured by a \emph{local} effective Hamilton function, amenable to Floquet engineering ~\cite{Higash_18}, even in chaotic classical many-body systems.
		
Even though Floquet theory does not apply to nonlinear EOM, a Magnus expansion (ME) can be formally defined for classical systems by replacing commutators with Poisson brackets~\cite{bukov2015universal,dalessio2013many}. The ME approximates the exact Floquet Hamiltonian $H_F\!\approx\! H_F^{(0+\dots+m)}\!\sim\!\mathcal{O}(\Omega^{-m})$, to a given order $m$ in the inverse frequency~\cite{supplementary}. However, it is an open question if and why the ME should work for classical systems. On one hand stands the notable application of the ME to the Kapitza pendulum~\cite{bukov2015universal,dalessio2013many,citro2015dynamical}, on the other -- the recent finding that the ME does not capture resonances, which renders its convergence at most asymptotic~\cite{bukov_15_res,weinberg2017adiabatic,claeys2017spin}. 

While energy is the most natural observable to study heating, it typically cannot be measured directly in experiments. We now show that the prethermal plateau affects generic local observables. To test Floquet theory, we initialize the system in the GS of $H_F^{(0+\dots+m)}$ with different values of $m$. Consider the staggered magnetization and its long-time average
\begin{equation*}
	M^{\alpha}(\ell T)\!=\!\frac{1}{N}\sum_{j=1}^{N} (-1)^{j} S_{j}^{\alpha}(\ell T);
	\overline{M^{\alpha}}\!=\!\frac{1}{10^3}\!\sum_{\ell=N_T}^{N_T+10^3}\!M^{\alpha}(\ell T),
\end{equation*}
where $N_T$ denotes a large number of driving cycles. This observable is an order parameter for AFM correlations, and its time dependence measures how well the system retains the initial AFM structure. Figure~\ref{fig:stgd_mag} (circles) shows the components of the time-averaged magnetisation $\overline{M^{\alpha}}$ as a function of frequency to order $m\!=\!0$. For $\Omega\lesssim \Omega_\ast$}, where $\Omega_\ast$ is the crossover frequency, the system enters quickly the infinite-temperature stage, and all information about the initial state is lost: $\overline{M^{\alpha}}=0$. However, in the long-lived prethermal plateau $\Omega\gtrsim \Omega_\ast$, much of the AFM correlations are preserved. The corresponding dashed lines show the staggered magnetization of the initial state, which is approached in the limit $\Omega/J\to\infty$. We verified that a similar behaviour is displayed by the spin-spin correlation function.

This raises the question whether one can Floquet-engineer expectation values of observables in the prethermal plateau. Upon increasing the order $m$ of the ME, we find a significant improvement between the staggered magnetization of the initial ensemble and its time-averaged value in the pre-thermal plateau (squares, triangles), cf.~Fig.~\ref{fig:stgd_mag}. We also checked that starting from the GS of $H_F^{(0)}$, and evolving the system with the time-independent $H_F^{(0+\dots+m)}$, results in better agreement of the magnetization dynamics with the exact stroboscopic evolution, upon increasing $m$~\cite{supplementary}. Since the ME is the main theoretical tool used in Floquet engineering~\cite{goldman2014periodically,bukov2015universal,eckardt2017atomic}, this result implies that one should be able to successfully Floquet-engineer the behavior of observables in the prethermal plateau in classical systems ~\cite{Higash_18}. Such a behavior likely originates from the emergent quasiconserved local integrals of motion for $\Omega\gtrsim \Omega_\ast$. 
	
\begin{figure}[t!]
	\centering
	\includegraphics[width=1.03\columnwidth]{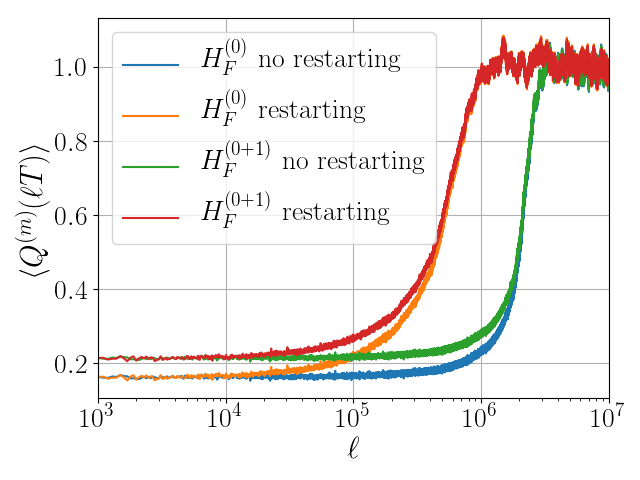}
	\caption{Heating behavior measuring $H_{F}^{(0)}$ and $H_{F}^{(0+1)}$ with and without ``restarting", for $\Omega/J \!=\! 3.8$. The restarting procedure is repeated every $10^3$ cycles [see text]. We define $\langle Q^{(m)}(\ell T)\rangle$ to measure the normalized excess energy w.r.t.~$H_F^{(0+\dots+m)}$, cf.~Eq.~\eqref{eq:Q_ave}. The difference in energy and its variance between the new and old ensembles was chosen to be less than $0.005J$. See Fig.~\ref{fig:stgd_mag} for parameters.}
	\label{fig:restarts}
\end{figure}
	
Floquet prethermal plateaus are defined with respect to the local approximate extensive Floquet Hamiltonian $H_F^{(0+\dots+m)}$. They are often assumed to be featureless states which are stroboscopically equivalent to thermal equilibrium with respect to $H_F^{(0+\dots+m)}$ for some optimal $m$. We now show that this assumption is incomplete: the prethermal plateau is sustained by drive-induced synchronization which is responsible for their exceptionally long stability. To demonstrate this, we compare the exact Floquet evolution, with an evolution where we repeatedly
restart the dynamics from the thermal Gaussian ensemble of $H_F^{(0)}$, with mean energy and width chosen to match those of the time-evolved initial ensemble into the prethermal plateau. We call this procedure ``restarting". Figure~\ref{fig:restarts} shows a comparison of the expected values of $H_F^{(0)}$, following the uninterrupted (blue) and ``restarting" (green) evolution. The ``restarting" procedure was applied every $10^3$ cycles to prevent the system from re-synchronizing. As a result, the ``restarted" dynamics enters the infinite-temperature stage exponentially earlier. 
The increase of energy caused by ``restarting" suggests that the prethermal plateau contains additional very slow synchronized dynamics, probably related to Arnold diffusion or subdiffusion~\cite{konishi1990diffusion,rajak_18,notarnicola2017localization}, which serves as a glue for the prethermal state. 
To argue that the prethermal plateau is a property of the time-evolved ensemble and not of the Gaussian energy ensemble used for ``restarting", we fix the initial ensemble of noisy AFM states based on the GS of $H_\mathrm{ave}=H_F^{(0)}$, and repeat the procedure measuring $H_F^{(0+1)}$, c.f.~Fig~\ref{fig:restarts}. As expected, this affects the energy density of the prethermal plateau but not the duration of the stage. 
The restarting procedure is carried out w.r.t.~a microcanonical distribution;  we checked both a microcanonical and canonical distribution and found the same results.
This restarting procedure shows that the pre thermal distribution \underline{cannot} be characterized by energy and energy variance alone. Specifically, if this were the case then the thermilzation times for the restarting and true evolutions would be the same. This restarting procedure is analogous to dephasing of density matrix in quantum mechanics.
	
	
\emph{Outlook.---}We revealed a dynamical heating transition between a prethermal and an infinite-temperature stage in the limit of infinite times and drive frequencies. Its existence influences strongly the evolution of periodically-driven many-body spin chains even at the experimentally-relevant moderate frequencies and times. This becomes manifest in a long-lived prethermal plateau, which can be modeled in a controllable fashion by an approximate effective nonlinear Floquet-Hamilton function derived within the limitations of the inverse-frequency expansion. Contrary to na\"ive expectations, the prethermal plateau is fused by drive-induced synchronization, and is not a featureless thermal state.
	
Even though a detailed comparison of thermalization in classical and quantum Floquet systems would be desirable, our analysis already presents compelling evidence that the prethermal plateau, observed in a variety of quantum models, survives in the classical limit~\cite{dalessio2013many,rajak_18}. This suggests that studies in cold atomic Floquet systems aiming to explain the contribution to heating due to higher bands or preparation of states under periodic driving, can be done (semi-)classically to reduce finite-size effects. In fact, our study directly relates to experimental platforms, such as shaken superfluid ultracold gases, where the physics is governed by a classical spin model~\cite{struck_13}, or photonic topological insulators~\cite{rechtsman_13,hafezi_14,mittal_14}, described by the nonlinear wave equation.
	
\emph{Note added:} While this manuscript was under peer review, a related work proving prethermalization in classical periodically-driven spin systems appeared~\cite{mori2018floquet}.

\emph{Acknowledgements.---}
We thank Pankaj Mehta and Takashi Mori for interesting and insightful discussions.
OH was supported by BU UROP student funding and the Simmons Foundation Investigator grant MMLS to Pankaj Mehta.
DS acknowledges support from the FWO as post-doctoral fellow of the Research Foundation -- Flanders and CMTV.
AP and PW were supported by NSF DMR-1813499, AFOSR FA9550-16-1-0334 and ARO W911NF1410540.
MB acknowledges support from the Emergent Phenomena in Quantum Systems initiative (EPiQS) of the Gordon and Betty Moore Foundation and ERC synergy grant UQUAM.
We used \href{https://github.com/weinbe58/QuSpin#quspin}{QuSpin} for simulating the nonlinear EOM in the case of monochromatic drive~\cite{weinberg_17,quspin2}. The authors are pleased to acknowledge that the computational work reported on in this paper was performed on the Shared Computing Cluster which is administered by \href{https://www.bu.edu/tech/support/research/}{Boston University's Research Computing Services}. The authors also acknowledge the Research Computing Services group for providing consulting support which has contributed to the results reported within this paper.

\bibliographystyle{apsrev4-1}
\bibliography{./Fclassical.bib}
\graphicspath{
	{allpaperfigs/},
}

\begin{widetext}

	\section*{\large Supplemental Material}

	\section{\label{sec:Ising}Analogy of the Physics of the Dynamical Heating Transition with the 1D Ising Model close to Zero Temperature}
	
	The thermalization dynamics of the dimensionless energy $\langle Q(\ell T)\rangle$ and energy variance $\langle \delta Q(\ell T)\rangle$ studied in this work are very similar in nature to the behavior of the squared magnetization in the one-dimensional Ising model close to zero temperature. Consider the 1d Ising energy function
	\begin{equation}
	\mathcal{H} = -J\sum_{j=1}^N \sigma_{j+1}\sigma_j,
	\end{equation} 
	where $\sigma_j\in\{\pm 1 \}$ are Ising variables and $J$ is the interaction strength. This model is a textbook example for a continuous phase transition between a disordered phase at finite temperature and a magnetized point at zero temperature. Below, we discuss how this critical point affects the small but finite temperature physics, and draw an analogy to the dynamical heating phase transition reported on in the main text.
	
	\begin{figure}[h!]
		\centering
		\begin{tabular}{ll}
			\includegraphics[width=0.5\columnwidth]{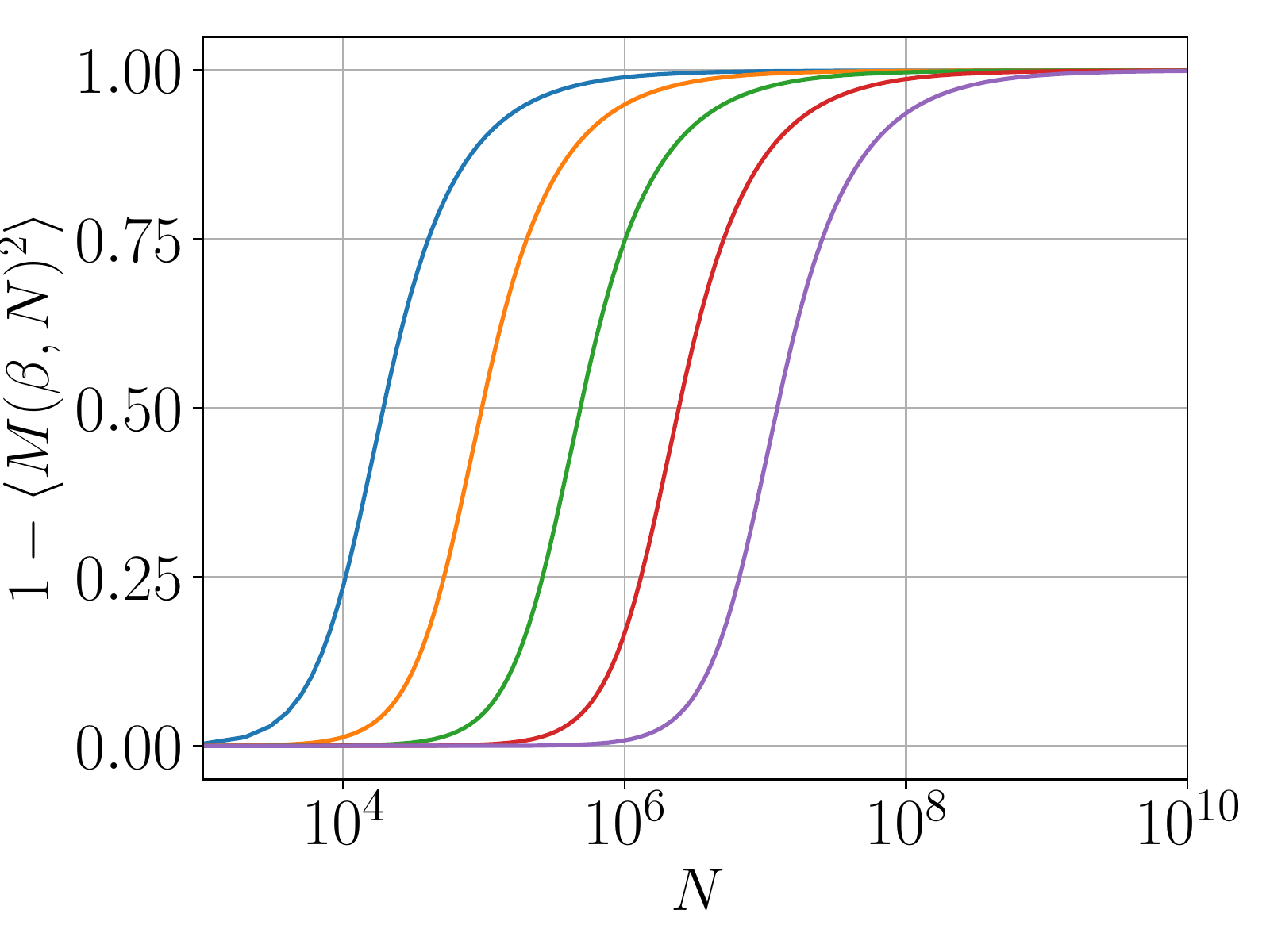}
			\includegraphics[width=0.5\columnwidth]{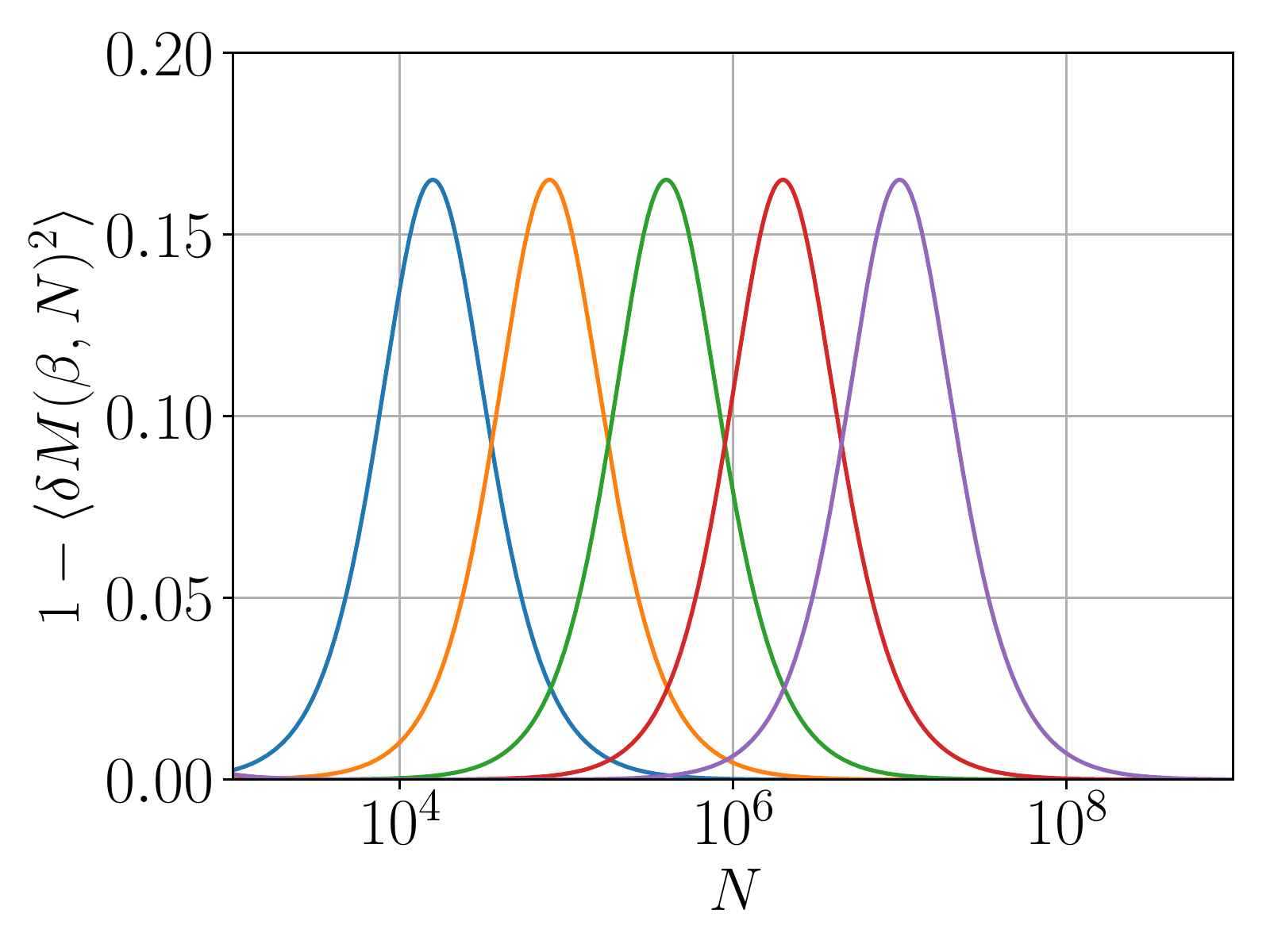}	
		\end{tabular}
		
		\begin{tabular}{ll}
			\includegraphics[width=0.5\columnwidth]{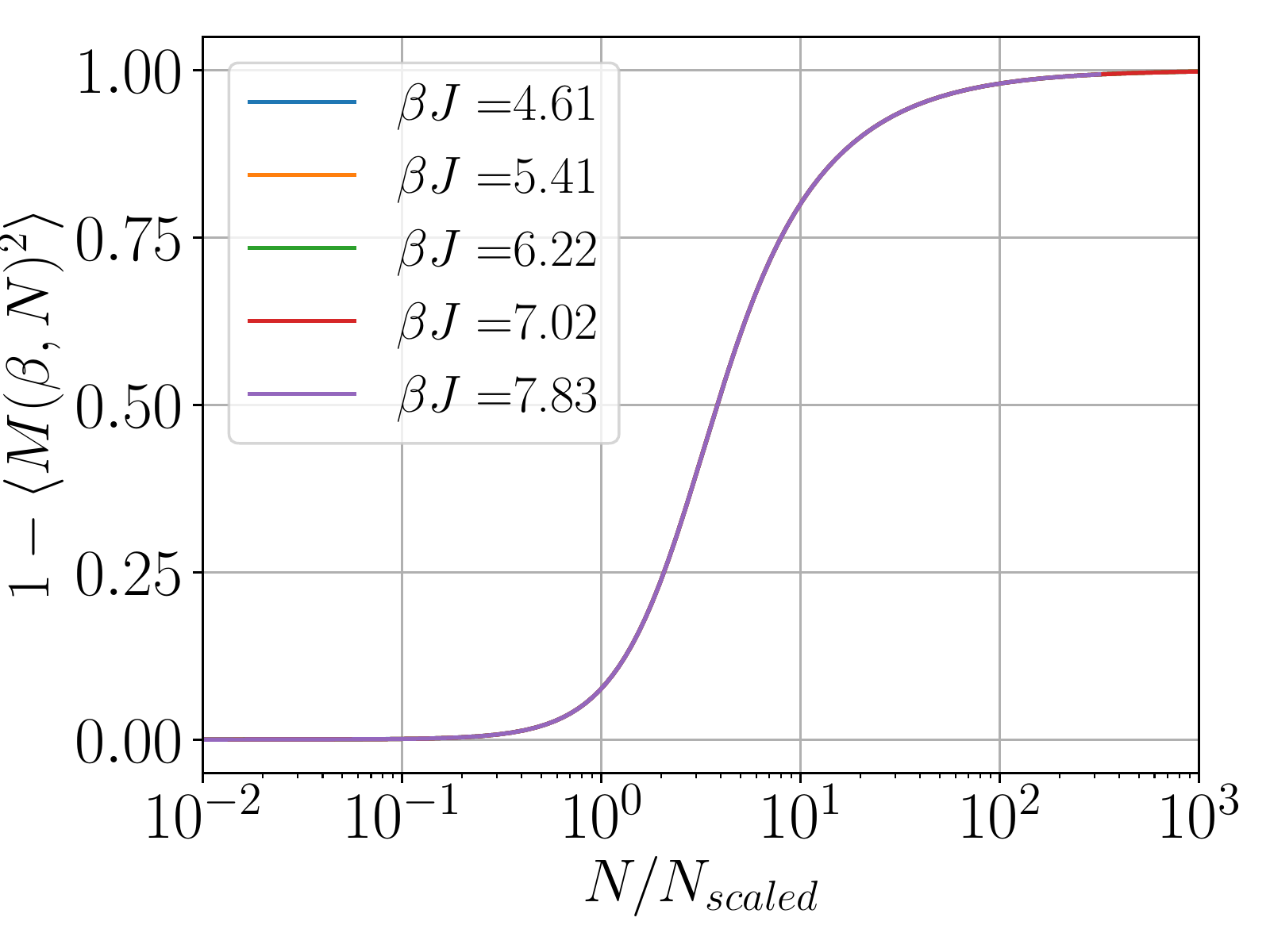}
			\includegraphics[width=0.5\columnwidth]{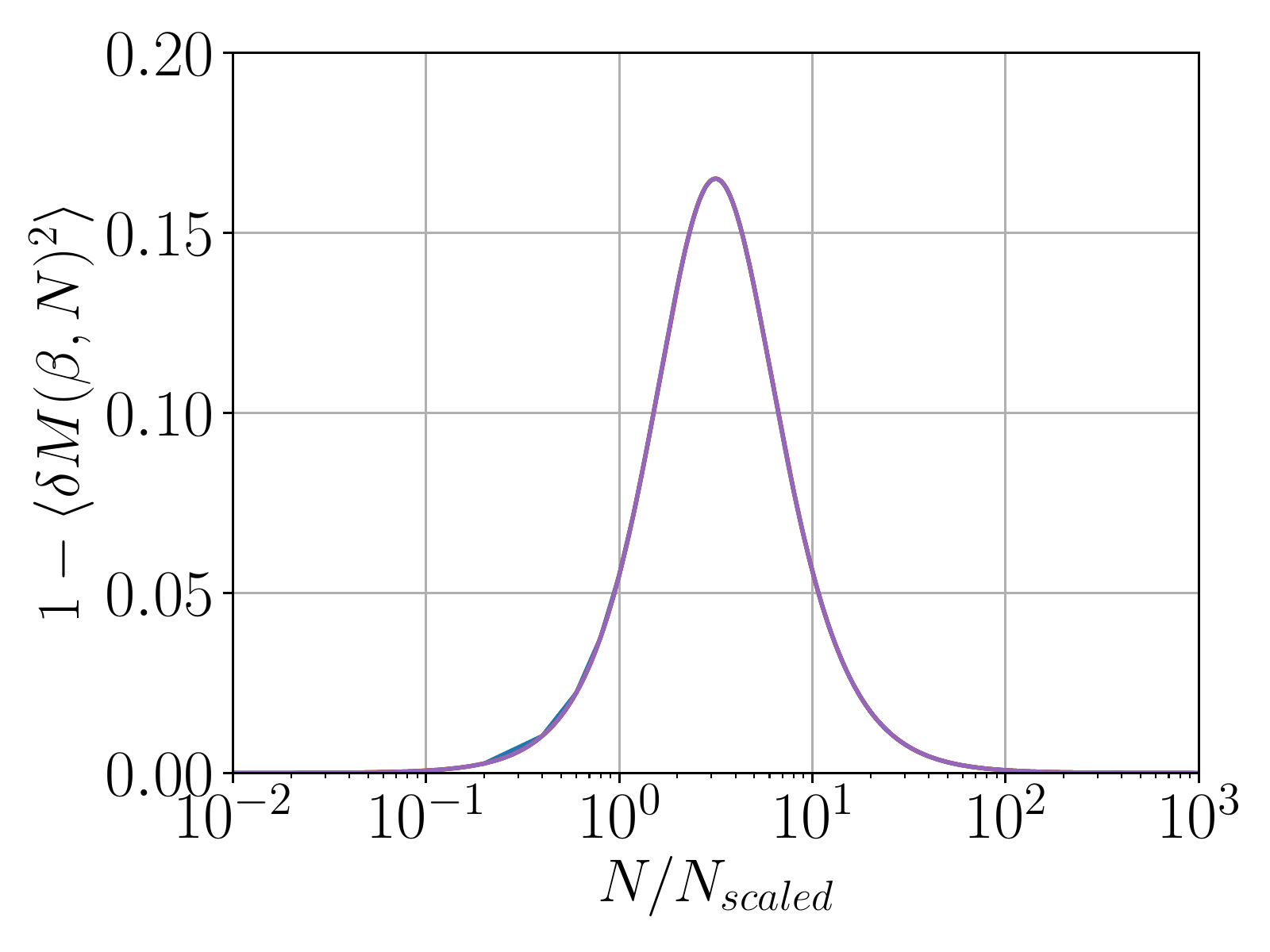}	
		\end{tabular}
		
		\caption{\label{fig:rescaleddM}The left plots show the behavior of the squared magnetization $\langle M^2(\beta,N) \rangle$ for different inverse temperatures $\beta$ as a function of the system size $N$. The right plots show the behavior of the variance of the squared magnetization $\langle  \delta M(\beta,N)^{2} \rangle$. The top panel on both sides shows the exact behavior while the bottom panel is rescaled with respect to the correlation length $N_\mathrm{scaled}=\xi(\beta)$ which is an exponential function of the inverse temperature. To be compared to Fig.~\ref{fig:rescaleddQ}.}
		
	\end{figure}

	\begin{figure}
		\centering
		\begin{tabular}{ll}
			\includegraphics[width=0.5\columnwidth]{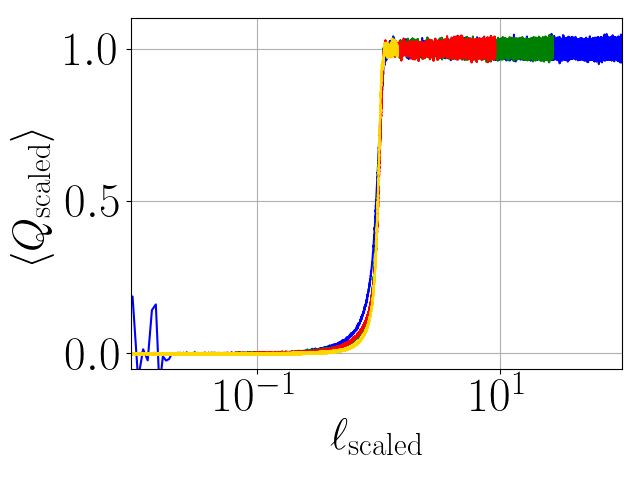}
			\includegraphics[width=0.5\columnwidth]{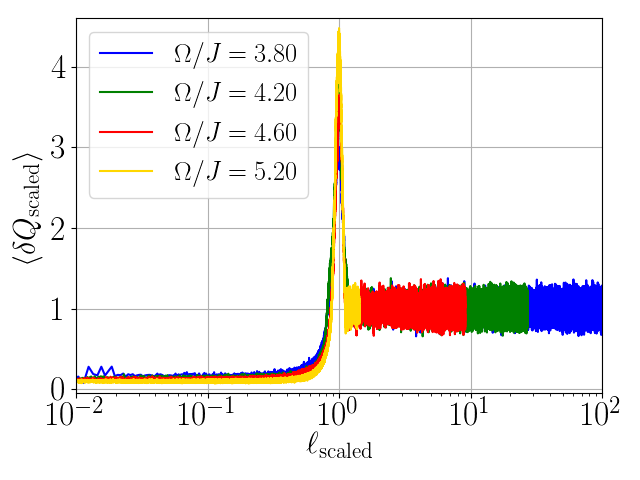}	
		\end{tabular}
		
		\caption{\label{fig:rescaleddQ}Noise-averaged energy $\langle Q( \ell T) \rangle$ and energy variance $\langle \delta Q( \ell T) \rangle$ as a function of the rescaled number of driving cycles $\ell_{\mathrm{scaled}}$. The simulation parameters are $h/J = 0.809$, $g/J = 0.9045$. To be compared to Fig.~\ref{fig:rescaleddM}.}
		
	\end{figure}
	
	Figure~\ref{fig:rescaleddM} shows the thermal expectation of the squared magnetization: $M^2=\left(N^{-1}\sum_i \sigma_i\right)^2$ and its variance as a function of the length of the chain $N$ for different inverse temperatures $\beta$. The curves show a crossover when the system size becomes of order the correlation length which goes as $\xi(\beta)\sim\exp(2J\beta)$.
	
	The scaling analysis for the heating transition described in the main text clearly exhibits the following scaling behaviour:
	\begin{equation}
	Q(lT,\Omega)=f_Q(l/l_\mathrm{max}(\Omega)),
	\end{equation}
	with $l_\mathrm{max}(\Omega)\sim\exp(c\Omega)$, see Fig.~\ref{fig:rescaleddQ}. This bares a striking resemblance to the equilibrium finite size scaling in the 1d Ising chain:
	\begin{equation}
	\langle M^2(\beta,N)\rangle = f_{M^2}(N/\xi(\beta)).
	\end{equation}
	Similar to defining a scaling time we can define a scaling length as $N_{\mathrm{scaled}} = \xi(\beta)$. This suggests that for the heating transition, $\Omega$ plays the role of $\beta$. At the same time, the correlation length, $\xi(\beta)=N_{\mathrm{scaled}}$, is analogous to the time needed to begin the unconstrained thermalization process. Furthermore this analogy gives a very strong evidence for a dynamical heating transition at $\Omega, t\rightarrow\infty$ as both the time dependent model and the equilibrium model share the same scaling behavior.

	\section{Time Evolution of the Staggered Magnetization}
	
	Figure~\ref{fig:test} shows the time-evolution of the magnetization for different frequencies, starting from the ground state of the period-averaged Hamiltonian $H_\mathrm{ave}$. Each data point is averaged over $50$ noise realizations of the initial state. For low frequencies the mean magnetization exhibits prethermal plateaus that eventually decay to zero. This decay originates from the breakdown of the anti-ferromagnetic structure of the initial state. 
	
	Figure~\ref{fig:test2} shows the behavior of the staggered magnetization for a single realization. In this case the magnetization exhibits spin flips which appear sharp even on a linear time scale.  These processes reveal the physical mechanism behind the onset of heating, which is associated with the proliferation of spin flips towards the end of the exponentially long-lived prethermal plateau. For moderate-to-high frequencies these magnetization plateaus persist for the entire duration of the prethermal plateau.

	\begin{figure}[t!]
		
		\begin{tabular}{ll}
			\includegraphics[width=0.333\columnwidth]{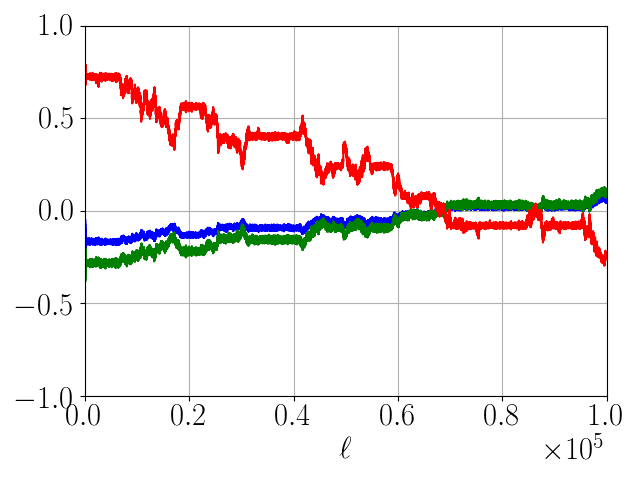}
			\includegraphics[width=0.333\columnwidth]{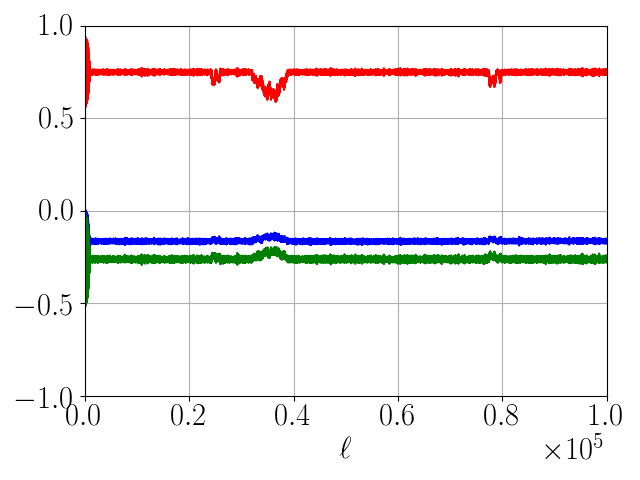}
			\includegraphics[width=0.333\columnwidth]{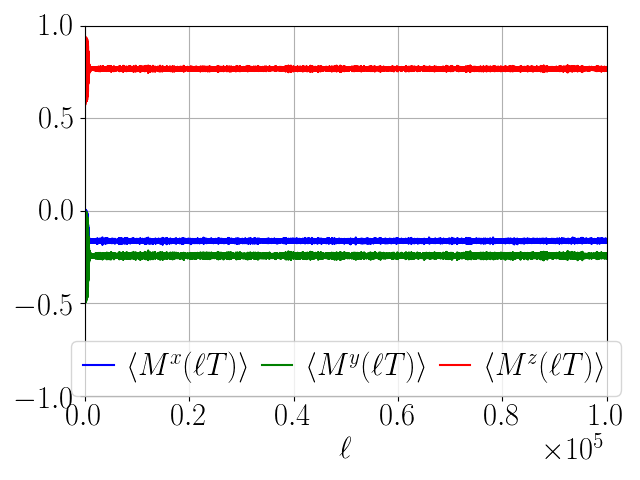}
			
		\end{tabular}
		
		\caption{The components of the staggered magnetization $\langle M^\alpha(\ell T)\rangle$ as a function of the number of driving cycles $\ell$, for $\Omega/J = 3.8$ (left), $\Omega/J=4.2$ (middle) and $\Omega/J=4.6$ (right). Each data point is averaged over $50$ noise-realizations. The simulation parameters are $h/J=0.809$, $g/J=0.9045$, and $N=100$.}
		\label{fig:test}
	\end{figure}
	
	\begin{figure}[t!]
		
		\begin{tabular}{ll}
			\includegraphics[width=0.333\columnwidth]{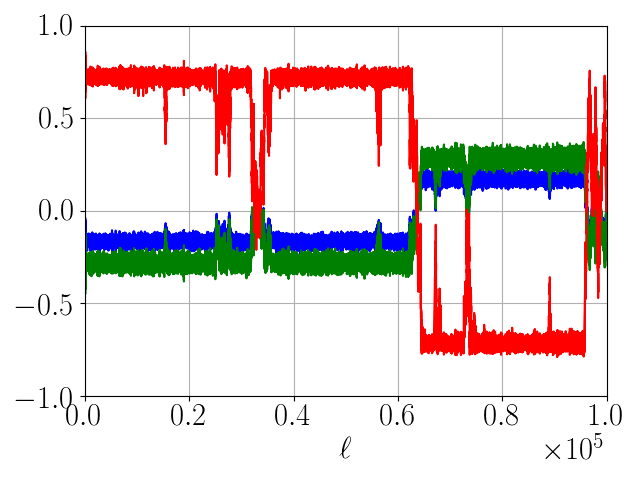}
			\includegraphics[width=0.333\columnwidth]{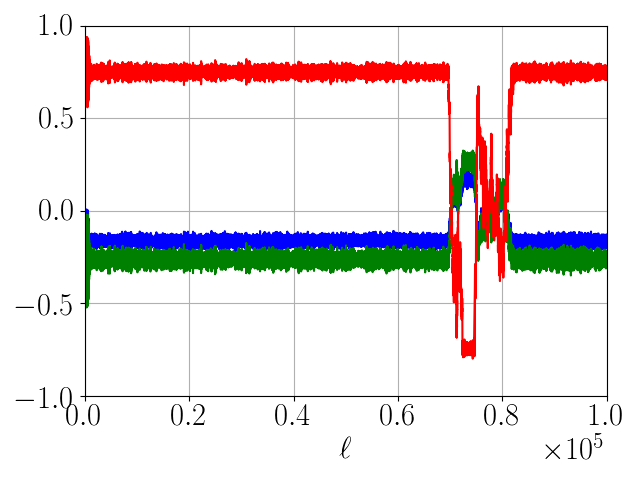}
			\includegraphics[width=0.333\columnwidth]{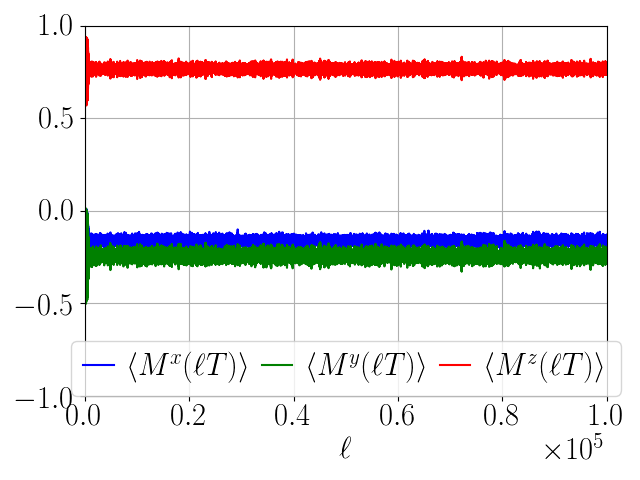}
			
		\end{tabular}
		
		\caption{A single noise realization showing the components of the staggered magnetization $\langle M^\alpha(\ell T)\rangle$ as a function of the number of driving cycles $\ell$, for $\Omega/J = 3.8$ (left), $\Omega/J=4.2$ (middle) and $\Omega/J=4.6$ (right). The simulation parameters are $h/J=0.809$, $g/J=0.9045$, and $N=100$.}
		\label{fig:test2}
	\end{figure}

	\section{\label{sec:heating}Frequency Dependence of Energy Absorption}
	
	Often times one is interested in the average amount of energy absorbed by the system from the drive. To measure this, let us define the following time-averaged energy absorption, averaged at a given decade over $10^3$ periods
	\begin{equation}
	\langle \overline{Q}^{N_T}(\Omega)\rangle  = \frac{1}{10^3}\sum_{\ell=N_T}^{N_T + 1000} \langle \overline{Q}(\ell T;\Omega)\rangle,
	\end{equation}
	where $N_T$ denotes the first driving cycle of the period used for the long-time average. Theoretically, one would like to do the period averaging at infinite time, $N_T\to \infty$. However, in classical systems this is not possible. Therefore, we study the heating dependence as a function of frequency, for a sequence of exponentially growing time spans. 
	
	The existence of a heating transition at $\Omega\to\infty$ and $\ell\to\infty$ suggests that the infinite-temperature plateau of $\langle \overline{Q}^{N_T}(\Omega)\rangle$ shifts to higher frequencies with increasing waiting time $N_T$. Figure~\ref{fig:decade} shows $\langle \overline{Q}^{N_T}(\Omega)\rangle$ over seven consecutive decades in $N_T$. In agreement with theory, as $N_T$ is increased, the energy curves shift gradually to the right. Interestingly, the crossover region between the infinite-temperature and the prethermal stages sharpens very slowly which suggests the existence of logarithmic corrections to the finite time scaling.

	\begin{figure}[t!]
		\centering
		\includegraphics[width=0.5\columnwidth]{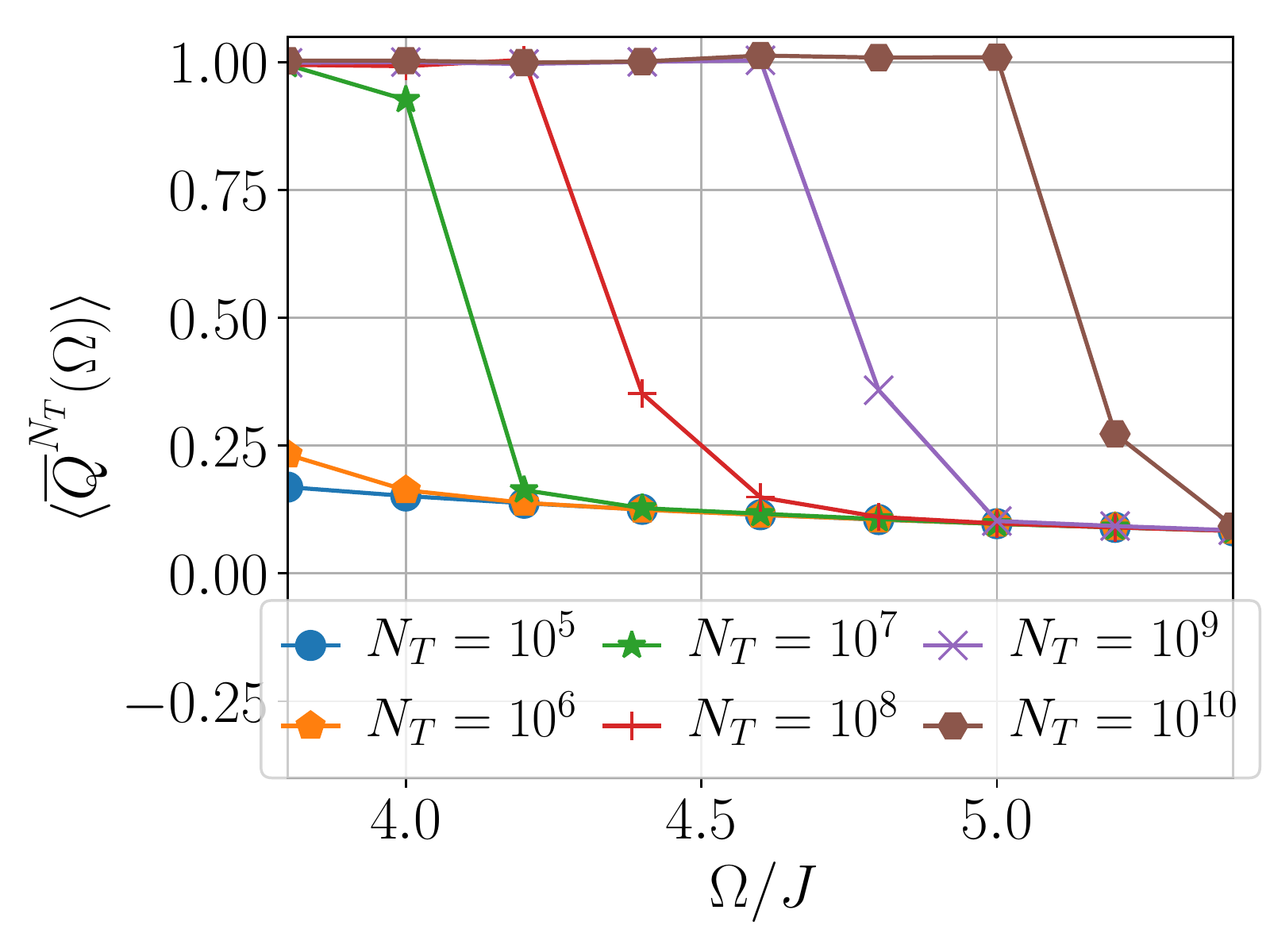}
		\caption{ Decade averaged energy absorption $\langle \overline{Q}^{N_T}(\Omega)\rangle$ for six different decades as a function of  the driving frequency $\Omega/J$. Each average is taken over 1000 periods. The simulation parameters are $h/J = 0.809$, $g/J = 0.9045$, and $N = 100$.}
		\label{fig:decade}
	\end{figure}

	\section{\label{sec:Magnus}Derivation of the Effective Floquet-Hamilton Function Using the Inverse-Frequency Expansion}
	
	The cornerstone of Floquet theory in physics is Floquet's theorem. However, the theorem applies to \emph{linear} ordinary differential equations, and hence its consequences become invalid for systems obeying nonlinear dynamics. On the other hand, chaos and thermalization are intimately tied to the nonlinearities in the Hamilton equations of motion. Therefore, finding effective descriptions for classical periodically-driven many-body systems is a challenging and difficult problem, and there might not exist a simple universal solution.
	
	In periodically-driven quantum systems, the linearity of the Schr\"odinger equation fulfils the critera for the applicability of Floquet's theorem. According to the latter, the evolution operator $U(t,0)$ factorizes as
	\begin{equation}
	\hat U(t,0) = \mathcal{T}_t\mathrm{exp}\left(-i \int_0^t\mathrm{d}t'\hat H(t')\right) = \hat P(t)\;\mathrm e^{-i t \hat H_F},
	\label{eq:Floquet_thm}
	\end{equation}
	with the so-called time-periodic micromotion operator $\hat P(t)=\hat P(t+T)$, which governs the fast intraperiod dynamics, and the effective time-independent Floquet Hamiltonian $\hat H_F$. Stroboscopically, the time evolution of the system is therefore described by the Floquet Hamiltonian $\hat H_F$.
	
	The starting point in quantum Floquet theory is the assumption that the Floquet Hamiltonian can be expanded in a Taylor series in the inverse frequency
	\begin{equation}
	H_F = \sum_{n=0}^\infty H_F^{(n)},\qquad H_F^{(n)}\sim \mathcal{O}(\Omega^{-n}), \qquad H_F^{(0+\dots+m)} = \sum_{n=0}^m H_F^{(n)}.
	\end{equation}
	Combining this with Eq.~\eqref{eq:Floquet_thm}, one arrives at the inverse-frequency expansion for the Floquet Hamiltonian, for more details see Refs.~\cite{goldman2014periodically,eckardt2017atomic,bukov2015universal}. A notable advantage of this expansion is that it remains valid beyond the linear response regime, and allows to study strongly driven systems.
	\begin{figure}[t!]
		
		\begin{tabular}{ll}
			\includegraphics[width=0.5\columnwidth]{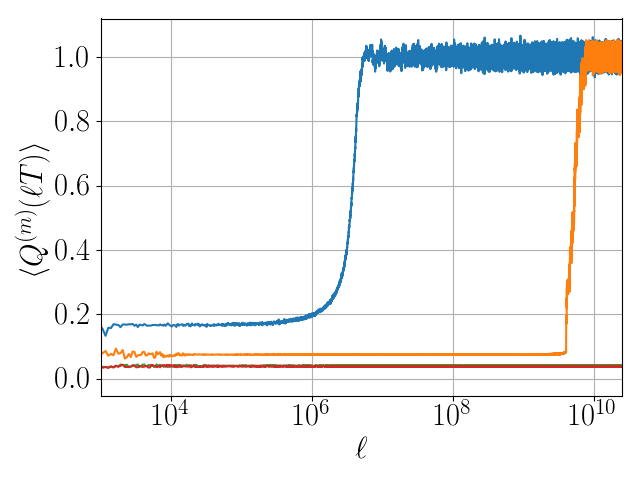}
			
			\includegraphics[width=0.5\columnwidth]{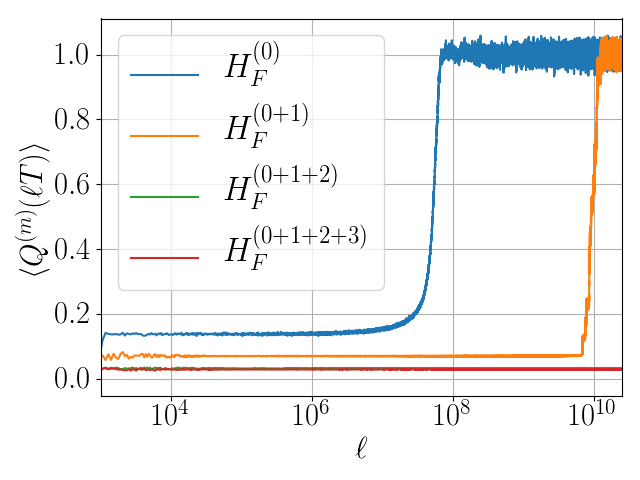}
		\end{tabular}
		\caption{Average energy absorption $\langle Q(t)\rangle$ as a function of the number of driving cycles $\ell$, for different $H_{F}^{(0+\dots+m)}$ for $m \in [0,1,2,3]$, see Eq.~\eqref{eq:Q_ME}. Higher order terms absorb less energy after the initial quench and exhibit less fluctuations in the prethermal plateau. The evolution is started in the ground state of $H_{F}^{(0+\dots+m)}$. The frequency values are $\Omega/J = 3.8$ (left) and $\Omega/J = 4.6$ (right). The simulation parameters are $h/J = 0.809$, $g/J = 0.9045$, and $N = 100$. }
		\centering
		\label{fig:longFloquet}
	\end{figure}

	\begin{figure}[t!]
		
		\begin{tabular}{ll}
			\includegraphics[width=0.5\columnwidth]{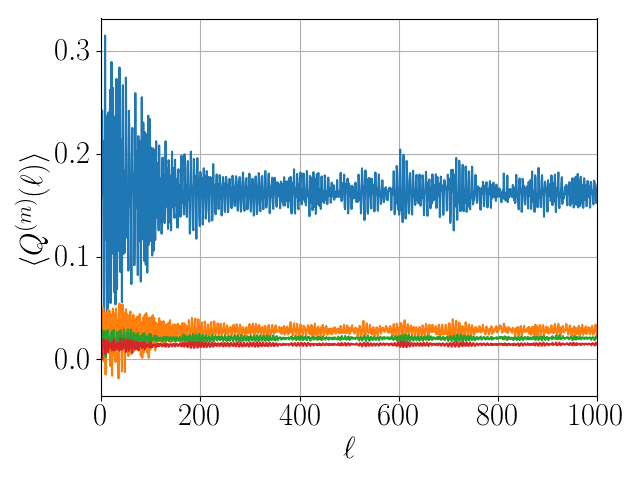}
			\includegraphics[width=0.5\columnwidth]{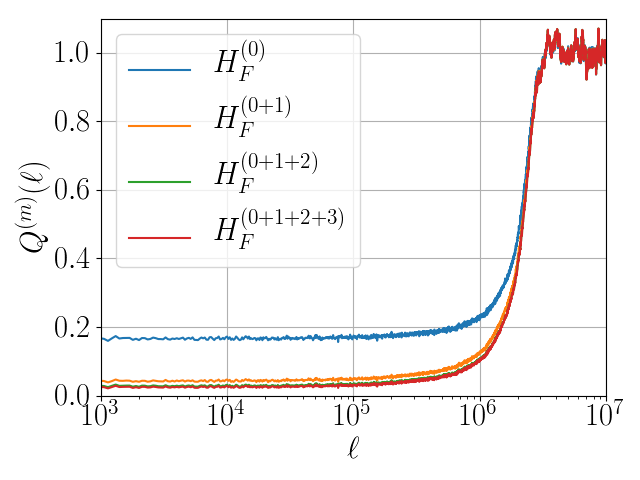}
		\end{tabular}
		\caption{Average energy absorption $\langle Q(t) \rangle$ as a function of the number of driving cycles $\ell$, for different $H_{F}^{(0+\dots+ m)}$ for $m \in [0,1,2,3]$. The evolution is started in the ground state of $H_{F}^{(0)}$ for all curves. The frequency value is $\Omega/J = 3.8$. The simulation parameters are $h/J = 0.809$, $g/J = 0.9045$, and $N = 100$. Note that when started in $H_{F}^{(0)}$ all Floquet expansions thermalize after the same number of periods but the evolution of higher order Floquet terms exhibit less fluctuations after the initial quench and in the prethermal plateau.  }
		\centering
		\label{fig:longFloquetH0gs}
	\end{figure}
	
	Let us apply the Magnus expansion -- one of the variants of the inverse-frequency expansion -- to the step-driven spin chain. Note that if we write the Hamiltonian as
	\begin{equation}
	\hat H(t) = 
	\begin{cases}
	\hat A & \text{for}\ t\in [0,T/2]\ \mathrm{mod}\ T  \\
	\hat B  & \text{for}\ t\in [T/2,T]\ \mathrm{mod}\ T 
	\end{cases}
	\nonumber
	\end{equation}
	where
	\begin{equation}
	\label{eq:AB}
	\hat A=\sum_{j=1}^{N} J \hat S_{j}^{z} \hat S_{j + 1}^{z} + h \hat S_{j}^{z},\qquad \hat B=\sum_{j=1}^{N} g \hat S_{j}^{x},
	\end{equation}
	the time evolution operator takes the form
	\begin{equation}
	\hat{U}(T,0) = \exp(-i \hat H_{F} T )=   \exp\left(-i \hat{B} \frac{T}{2}\right) \exp\left(-i \hat{A}\frac{ T}{2}\right).
	\end{equation}
	For this setup, applying the Magnus expansion is equivalent to the Baker-Campell-Hausdorff formula, which to third order in the inverse frequency gives:
	\begin{eqnarray}
	\hat H_\mathrm{ave}=\hat H_{F}^{(0)} &=& \frac{1}{2} (\hat A + \hat B) ,\nonumber \\
	\hat H_{F}^{(1)} &=& \frac{T}{8}[\hat B,\hat A] ,\nonumber\\
	\hat H_{F}^{(2)} &=& \frac{T^{2}}{96} \left( [\hat B,[\hat B,\hat A]] + [\hat A,[\hat A,\hat B]] \right) ,\nonumber \\
	\hat H_{F}^{(3)} &=& -\frac{T^{3}}{284} [\hat A,[\hat B,[\hat B,\hat A]]].
	\end{eqnarray}
	In Ref.~\cite{bukov2015universal} it was noted that if we formally replace the commutator $[\cdot,\cdot]$ by the Poisson bracket $\{\cdot,\cdot\}$, the expansion translates over to classical systems. In our model, this leads to
	\begin{eqnarray}
	H_{F}^{(0)} &=& \frac{1}{2}\sum_{j=1}^{N} J S_{j}^{z} S_{j+1}^{z} + h S_{j}^{z} + g S_{j}^{x} ,\nonumber\\
	H_{F}^{(1)} &=& \frac{-gT}{8} \sum_{j=1}^{N} (J(S_{j+1}^{z} + S_{j-1}^{z})   + h)S_{i}^{y} ,\nonumber\\ 
	H_{F}^{(2)} &=& \frac{T^{2}}{96}  \sum_{j=1}^{N}  Jg^{2}( S^{y}_{j} ( S^{y}_{j+1} + S^{y}_{j-1} ) - S_{k}^{z}( S^{z}_{j+1} + S^{z}_{j-1}))  - g^{2}hS^{z}_{j} \nonumber\\
	&& -2Jgh ( S_{j+1}^{z} + S_{j-1}^{z} )S_{j}^{x} - gh^{2}S_{j}^{x} -J^{2}g(  2S_{j+1}^{z}S_{j-1}^{z} + (S_{j+1}^{z})^{2}  + (S_{j-1}^{z})^{2}   ),\nonumber \\
	H_{F}^{(3)} &=& \frac{T^{3}}{284} \sum_{j=1}^{N} Jg^{2}hS_{j}^{x}( S_{j+1}^{z} + S_{j-1}^{z} ) + J^{2}g^{2}S_{j+1}^{z}S_{j}^{x}( S_{j+1}^{z} + S_{j-1}^{z} ) + J^{2}g^{2}S_{j}^{z}S_{j+1}^{x}( S_{j+2}^{z} + S_{j}^{z} ).
	\end{eqnarray}
	The above approach is equivalent to starting from a classical periodically-driven system, quantizing it, applying the Magnus expansion, and then taking the classical limit. While this procedure is straightforward to apply, the absence of a Floquet theorem for classical chaotic systems casts doubt on how controlled this expansion is.
	
	To study this, for every order $m$ in the Magnus expansion, we start from the GS of the corresponding approximate classical Floquet-Hamilton function $H_F^{(0+\dots+m)}$. The exact GS of $H_F^{(0+\dots+m)}$ can be computed by first making use of the two-site translational invariance of the system to effectively reduce the number of degrees of freedom. Except for $m=0$, we have a nonzero $S_{j}^{y}$ component, induced by the $y$-field in the first-order Floquet-Hamilton function. To find the correct ground state we, therefore, parametrize the spin vector $\vec S_j$ by spherical coordinates $(\theta_j,\phi_j)$, $j=1,2$, and variationally determine the angles through minimizing the energy $E_\mathrm{GS}^{(0+\dots+m)}(\theta_j,\phi_j)$. The true ground state must thus be solved for numerically. Note that the ground state for $m>0$ is also dependent on the driving frequency $\Omega$. 
	
	We then evolve the initial state with the exact time-dependent Hamilton function $H(t)$, and measure the energy $H_F^{(0+\dots+m)}$. Figure~\ref{fig:longFloquet} shows a comparison between the different order Magnus expansion for energy absorption [see main text for a comparison on the staggered magnetization]. The normalized energy-absorption quantity $Q$ is more generally defined as:
	\begin{equation}
	\langle Q^{(m)}(\ell T) \rangle \!=\! \frac{ \langle H_F^{(0+\dots+m)}[\{\vec S_j(\ell T)\}] \rangle - E_\mathrm{GS}^{(0+\dots+m)} }{\langle H_F^{(0+\dots+m)}\rangle_{\beta = 0} - E_\mathrm{GS}^{(0+\dots+m)} }.
	\label{eq:Q_ME}
	\end{equation}	
	Starting from the GS of the Magnus-corrected Hamiltonian, we find that the higher-order terms have the following salient effects, see Fig.~\ref{fig:longFloquet}. Firstly, the initial energy fluctuations arising from the ferromagnetic structure of the ground state during the constrained themalization [stage (i) in Fig.~1, main text] are reduced. In addition, the unconstrained thermalization transient [stage (iii) in Fig.~1, main text] shows a sharper transition with increasing order of the expansion. 
	
	To study the effect of the initial state on the dynamics, we can fix the initial GS to be the GS of $H_F^{(0)}$, and repeat the procedure, measuring the energy functions $H_F^{(0+\dots+m)}$, see Fig.~\ref{fig:longFloquetH0gs}. Clearly, while the pre-thermal plateau is reduced with increasing the order of the expansion, the heating time now remains unchanged. However, starting the evolution in the GS of a higher order Floquet Hamiltonian increases the duration of the prethermal plateau. If the evolution was completely described by quench with respect to some high order Floquet Hamiltonian, then the corresponding energy function $H_F^{(0+\dots+m)}$ of that order should be approximately conserved except for some energy absorption when the drive is turned on. We observe that this is true in the prethermal plateau but the heating time is not changed. The fact that the heating time does not change means that the heating mechanism is not completely explained by the Magnus expansion and the heating time is a function of the initial state.
	
	One might wonder how well the approximate Floquet-Magnus Hamiltonians describe the dynamics of observables within the prethermal plateau. In order study this, we prepare the system in the GS of $H_F^{(0)}$, and evolve it with the time-independent approximate $H_{F}^{(0+\dots+m)}$. Our aim is to compare the approximate dynamics, generated by $H_{F}^{(0+\dots+m)}$, to the exact periodic-step evolution. To compare the two kinds of evolution, we look into the staggered $z$-magnetization, $\langle M^{z}(\ell)\rangle$, which serves as an order parameter for the anti-ferromagnetic correlations in the GS, as mentioned in the main text. Since it is difficult to compare rapidly oscillating observables over long times, we also look at the stroboscopic time-average of $M^{z}$:
	\begin{equation}
	\langle \overline{M}^{z} (\ell)\rangle = \frac{1}{\ell} \sum_{\kappa=0}^{\ell} \langle M^{z}(\kappa) \rangle.
	\label{eq:Mz_ave}
	\end{equation}
	We find that quenching with a higher-order Floquet Hamiltonian reproduces the exact dynamics with greater fidelity, cf.~Fig.~\ref{fig:q}. This shows that the dynamics following the exact evolution is better captured as the order of the expansion is increased, although this behavior might reverse at some higher order, as is common for asymptotic expansions~\cite{mori_15,abanin_15}. That is an intriguing result because there is no straightforward generalization of Floquet's theorem to nonlinear systems.

	\begin{figure}[t!]
		
		\begin{tabular}{ll}
			\includegraphics[width=0.5\linewidth]{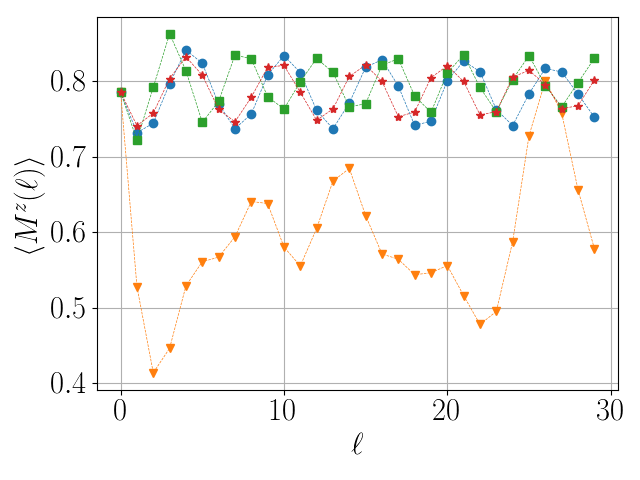}
			\includegraphics[width=0.5\linewidth]{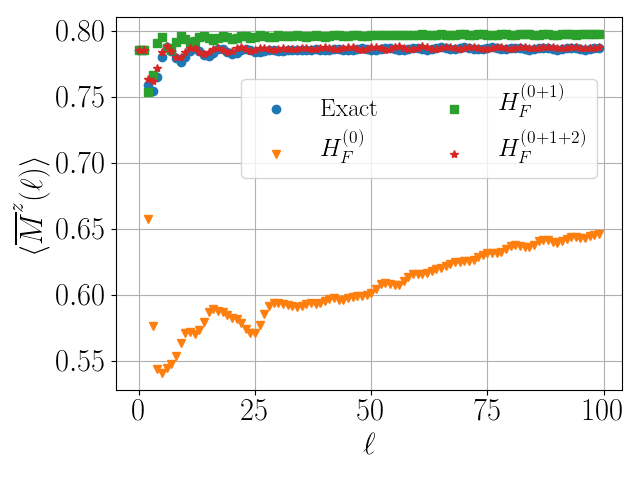}
			
		\end{tabular}
		
		\caption{Comparison of stroboscopic staggered magnetization $ \langle M^{z}(\ell) \rangle $ and average stroboscopic staggered magnetization $ \langle \overline{M}^{z}(\ell) \rangle $, see Eq.~\eqref{eq:Mz_ave}, as a function of the number of driving periods. The thin dashed lines are only a guide to the eye and were not used to do time averaging. The stroboscopic dynamics is generated by the exact step-drive evolution and different orders of the time-independent Floquet Hamiltonian, respectively, started from the same initial state. Note that going to higher order in the Magnus expansion increases the accuracy of the dynamics both stroboscopically (left panel) and in long-time average (right panel). The simulation parameters are $h/J = 0.809$, $g/J = 0.9045$, $N = 100$ and  $\Omega/J = 4.0$.}
		\label{fig:q}
	\end{figure}

	\section{Prethermal Phase}
	
	The behavior of the system in the prethemal phase looks like a thermal state for most local observables like magnetization. However, we argue that the prethermal phase is not completely described by a thermal distribution. This is shown in Figure 3 of the main text but we expand on this argument here.
	Consider the following numerical experiment: We take a ensemble of states in the prethermal regime and evolve with exact evolution for $1000$ periods and then evolve with $H_{F}^{(0)}$ or $H_{F}^{(0+1)}$  for $1000$ periods. Next, we take the same initial state and evolve with $H_{F}^{(0+1+2)}$ for $1000$ periods and quench with respect to $H_{F}^{(0)}$ or $H_{F}^{(0+1)}$ for $1000$ periods.

	\begin{figure}[t!]
		
		\begin{tabular}{ll}
						\includegraphics[width=0.5\columnwidth]{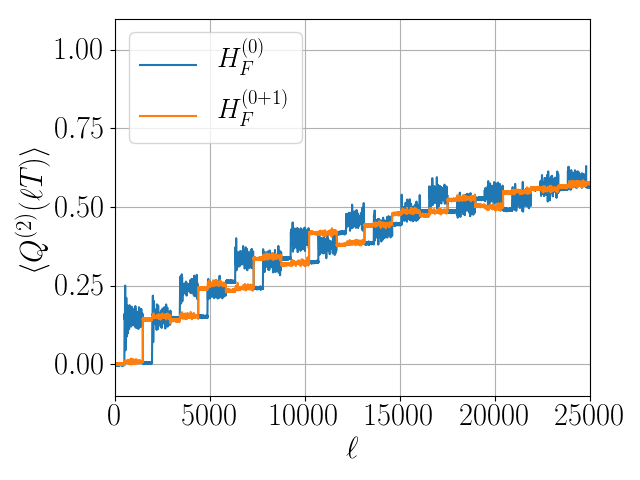}	
			\includegraphics[width=0.5\columnwidth]{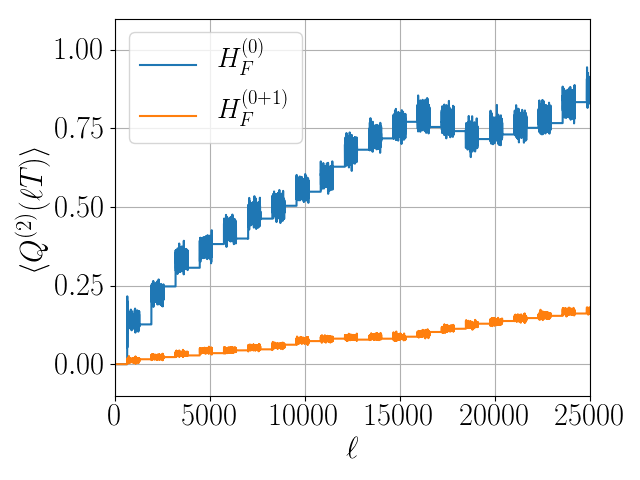}

		\end{tabular}
		
		\caption{ The left panel shows average energy absorption $\langle Q^{(2)}(t) \rangle$ as a function of the number of driving cycles $\ell$ for dynamics with alternating quenches between $H_{F}^{(0)}$ and exact evolution (left panel blue curve) and $H_{F}^{(0+1)}$  and exact evolution (left panel orange curve). The right panel shows average energy absorption $\langle Q^{(2)}(t) \rangle$ as a function of the number of driving cycles $\ell$ for dynamics with alternating quenches between $H_{F}^{(0)}$ and $H_{F}^{(0+1+2)}$ evolution (right panel blue curve) and $H_{F}^{(0+1)}$  and $H_{F}^{(0+1+2)}$ evolution (right panel orange curve). The length of each quench is 1000 periods. The graphs shown have $\Omega/J = 4.0$. The simulation parameters are $h/J = 0.809$, $g/J = 0.9045$, and $N = 100$. }
		\centering
		\label{fig:quenching}
	\end{figure}

	Figure~\ref{fig:quenching} shows that when evolving with $H_{F}^{(0+1+2)}$ the heating rate of quenching with respect to $H_{F}^{(0+1)}$ is much less then that of $H_{F}^{(0)}$. When we evolve with the exact evolution we find that the rates of heating of $H_{F}^{(0+1)}$ and $H_{F}^{(0)}$ are almost identical. This implies that the description of the pre thermal phase is not completely captured by a thermal ensemble. The heating we see as a result of restarting is inconsistent with the picture that we simply quench to the wrong Hamiltonian because then we would observe a very strong difference of the heating rate for restarting with $H_{F}^{(0)}$ and $H_{F}^{(0+1)}$. This observation leads us to believe that behavior of system in the pre thermal phase is not completely described by a thermal state and may contain additional slow coherent dynamics that are destroyed by quenching or restarting processes.

	\section{Noisy Drives and Synchronization}
	In this paper we have considered only driving periods of constant duration. If small amounts of noise are added to the duration of each driving frequency $T$ then we find a much faster rate of heating. This suggests that the long lifetime of the prethemal regime as seen in Figure 1 in the main text may be due to the emergence of quasi-conserved quantities caused by synchronization effects.
	
	To test this, before each new driving cycle, we slightly change the cycle duration $T'$ nondeterministically. In particular, using the relation $\Omega'=2\pi/T'$, we draw a new drive frequency $\Omega' = \Omega + \mu$; $\mu$ is a random number drawn from the uniform probability distribution supported on the interval $[0, 2\sigma\sqrt{3}]$, and $\sigma$ is the variance of the distribution which sets the strength of the noise. Figure~\ref{fig:noisy_period} shows that the rate of heating is very susceptible to such noise. This suggests that the existence of the long-lived prethermal regime in the noiseless case is due to some form of coherence effects.

	\begin{figure}[t!]
		
		\begin{tabular}{ll}
			\includegraphics[width=0.333\columnwidth]{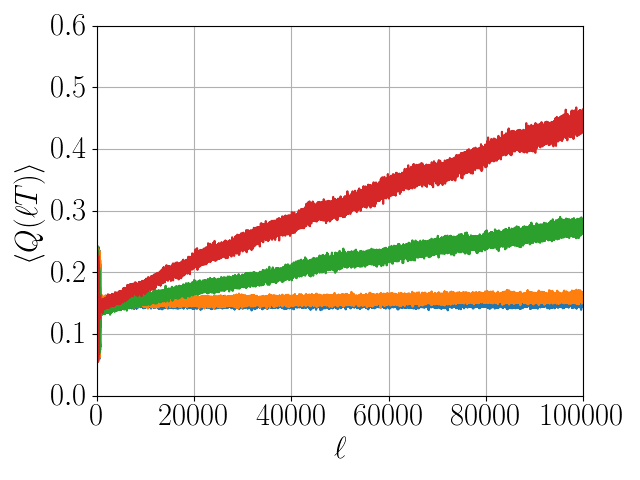}
			\includegraphics[width=0.333\columnwidth]{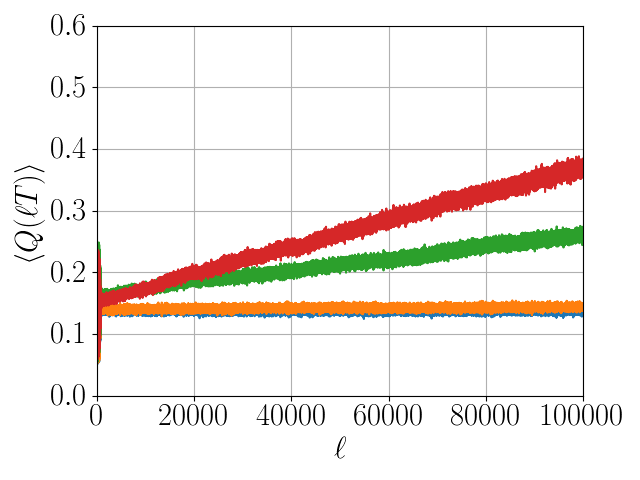}		\includegraphics[width=0.333\columnwidth]{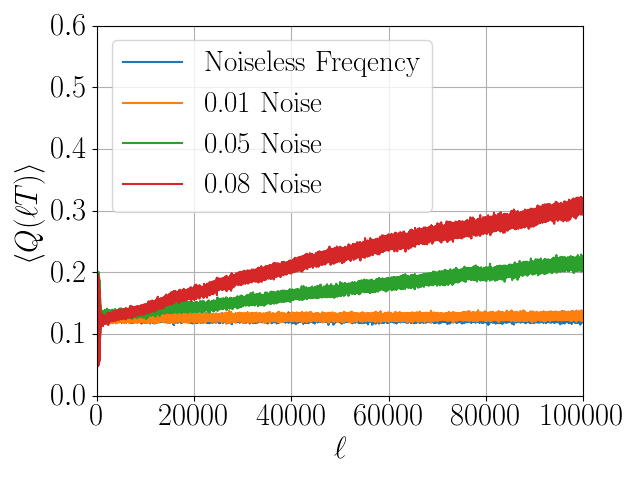}
			
		\end{tabular}
		
		\caption{ Average energy absorption $\langle Q(t) \rangle$ as a function of the number of driving cycles $\ell$ for different noisy driving durations [see text]. The evolution goes up to $10^{5}$ cycles. The graphs shown have $\Omega/J = 4.0$ (left), $\Omega/J = 4.2$ (middle) and $\Omega/J = 4.4$ (right). The simulation parameters are $h/J = 0.809$, $g/J = 0.9045$, and $N = 100$. The legend applies to all three graphs.  }
		\centering
		\label{fig:noisy_period}
	\end{figure}

	\section{Dependence of $g/J$ coupling on Thermalisation timescale }
	
	Our model is ergodic as long as the $g/J$ dimensionless parameter is non-zero, when $g/J = 0$ the model becomes the standard Ising model and is exactly solvable. The parameter $h/J$ is irrelevant for the physics of pre-thermalization since $H_{F}^{(0)}$ is non-integrable even for $h/J=0$ whenever the on-site spin degree of freedom $s>1/2$ [for $s=1/2$ one has the transverse-field Ising model which is integrable]. The right panel of Figure~\ref{fig:gjdependence} shows the heating time as a function of $g/J$ parameter when $h/J = 0$. Interestingly, the spacing between different $\Omega/J$ values for fixed $g/J$ is approximately constant on a log-log scale meaning that for $h/J = 0$ the heating time can be parametrized as $\ell_\mathrm{max}(\Omega)= r(g/J) \exp[-\gamma(g/J)\times\Omega/J]$, with $\gamma(g/J)$ a slowly-varying function of $g/J$, and $r(g/J)\propto (g/J)^{\alpha}$ with numerically determined exponent $\alpha \approx -2.12$. This would suggest that for fixed $g/J$ the exponential scaling (with the possibility of logarithmic corrections) is a \emph{universal} property. The left panel of Figure~\ref{fig:gjdependence} shows that for $g/J$ greater than $1.7$ the prethermal plateau is no longer present for low frequencies and the system begins to absorb energy immediately. This possibly explains why data for large $g/J$ the data does not have the same clean linear relationship on log-log scale as seen for low $g/J$ values. Simulating $g/J$ values less than $0.6$ is not feasible due to the divergence in heating time. This constrains the available window of meaningful data to $ 0.7 \le g/J \le 1.6 $

	\begin{figure}[t!]
		
		\begin{tabular}{ll}
			\includegraphics[width=0.5\linewidth]{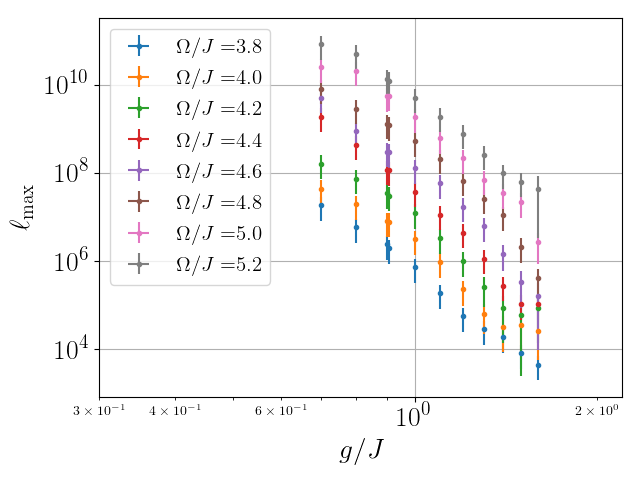}
			\includegraphics[width=0.5\linewidth]{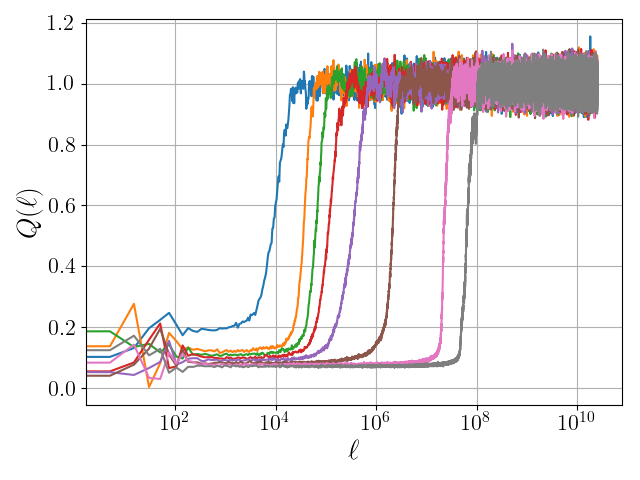}
			
		\end{tabular}

		\caption{ The left panel shows the heating time as a function of $g/J$ for different $\Omega/J$ values for $h/J=0$. The error bars are determined using a bootstrap method. There are $15$ realizations per data point. Note that the distance between two points of different frequency for the same $g$ value is approximately constant on logarithmic scale. The right panel shows the energy absorption as a function of time for $g/J = 1.6$. } 
		\label{fig:gjdependence}
		\centering
		
	\end{figure}

	\section{Mono- vs.~Polychromatic Drives. The Effect of Micromotion.}
	
	Throughout this study we used a time-periodic step drive function. This is a key property which ultimately allows to reach astronomically long times required to study pre-thermalization. However, typically in experiments it is easier to apply periodic modulations with few harmonics, while step drives require a more precise apparatus. One may, therefore, wonder how the heating properties of the system might change, if we used few-harmonic drives instead.
	
	Here, we give evidence that the form of the drive is relatively unimportant when considering the qualitative behavior of thermalizing dynamics. Let us cast the Hamilton function in the generalized form
	\begin{equation}
	H(t) = \frac{1}{2} (A + B)  + \frac{f(t)}{2} (A - B),
	\end{equation}
	see Eq.~\eqref{eq:AB}. For a step drive, we can Fourier-decompose the protocol as follows
	\begin{equation}
	f(t) 
	= \frac{4}{\pi} \sum_{n=1}^{\infty} \frac{1}{2n -1} \sin\left((2n-1)\Omega t  \right)
	=  
	\begin{cases} +1 & \text{for}\ t\in [0,T/2] \  \mathrm{mod}\ T   \\
	-1 & \text{for} \ t \in[T/2,T]  \ \mathrm{mod}\ T \\
	\end{cases}
	\end{equation}
	Note that each term of the expansion of $f(t)$ has zero average over a period. Thus, no matter the number of terms at which the series is truncated, $H_{F}^{(0)} = \frac{1}{2} ( A + B )$. 
	
	Figure~\ref{fig:driving_comp} shows a comparison of the energy absorption with $f(t)$ truncated at different orders in the Fourier expansion. The graphs show that the presence of higher order Fourier modes only marginally increases the rate of heating throughout the frequency range of interest.  Hence, the picture developed in the main text about the properties of the thermalizing dynamics remains robust to the number of harmonics the driving protocol is composed of. Furthermore, this graphs the behavior for all times not just stroboscopically. This suggests that irrespective of the phase of the drive (i.e.~the Floquet gauge), the qualitative behavior of dynamics remains the same. We used QuSpin~\cite{weinberg_17,quspin2} to simulate the nonlinear EOM.

	\begin{figure}[t!]
		
		\begin{tabular}{ll}
			\includegraphics[width=0.333\columnwidth]{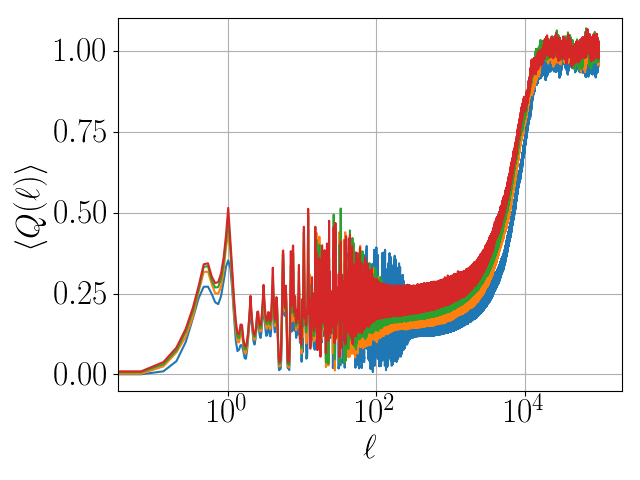}
			\includegraphics[width=0.333\columnwidth]{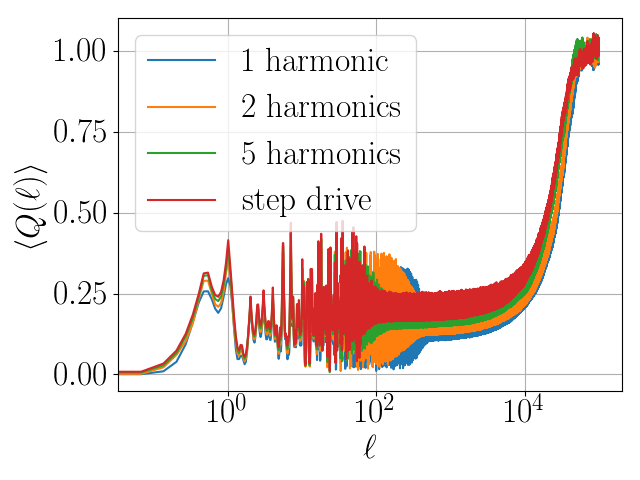}
			\includegraphics[width=0.333\columnwidth]{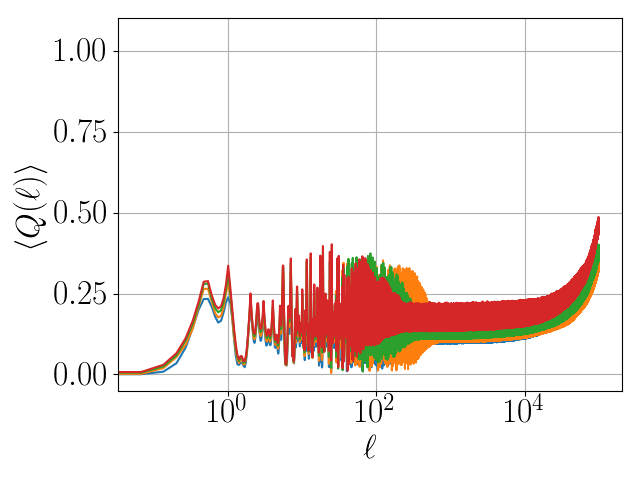}
		\end{tabular}
		
		\caption{Average energy absorption $\langle Q(t) \rangle$as a function of the number of driving cycles $\ell$ for different truncations of $f(t)$. The graphs shown have $\Omega/J = 3.0$ (left), $\Omega/J = 3.2$ (middle) and $\Omega/J = 3.4$ (right). The simulation parameters are $h/J = 0.809$, $g/J = 0.9045$, and $N = 100$.  This plot shows 15 points per period going beyond stroboscopic dynamics. }
		\centering
		\label{fig:driving_comp}
	\end{figure}

	\section{\label{sec:scaling}System Size Dependence}
	Figure~\ref{fig:size_dep} shows that the timescale at which the prethermal plateau ends and thermalisation occurs is insensitive to the system size. However, the steepness of this heating process does increase with increasing the system size. Additionally, in the infinite-temperature state [stage (iv), cf.~Fig.~1, main text] reaching larger system sizes reduces the size of the temporal fluctuations, as anticipated from the laws of Statistical Mechanics.

	\begin{figure}[t!]
		
		\begin{tabular}{ll}
			\includegraphics[width=0.333\columnwidth]{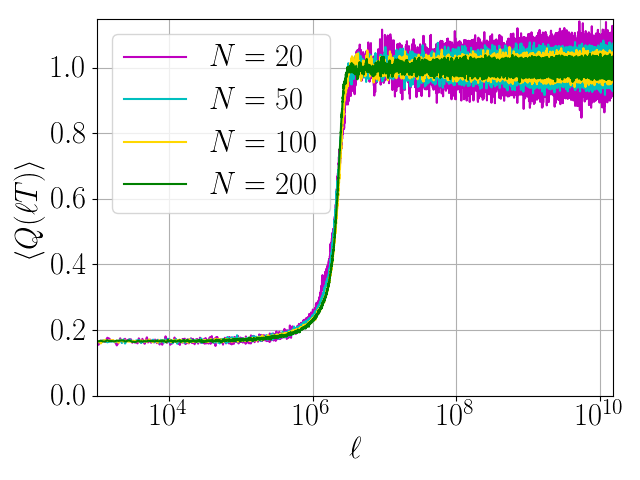}
			\includegraphics[width=0.333\columnwidth]{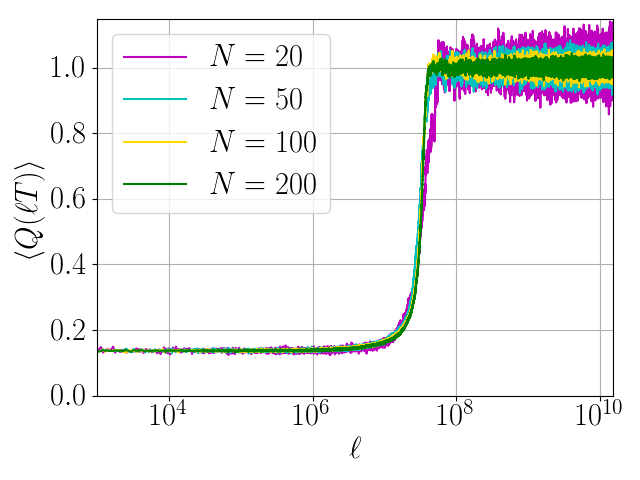}
			\includegraphics[width=0.333\columnwidth]{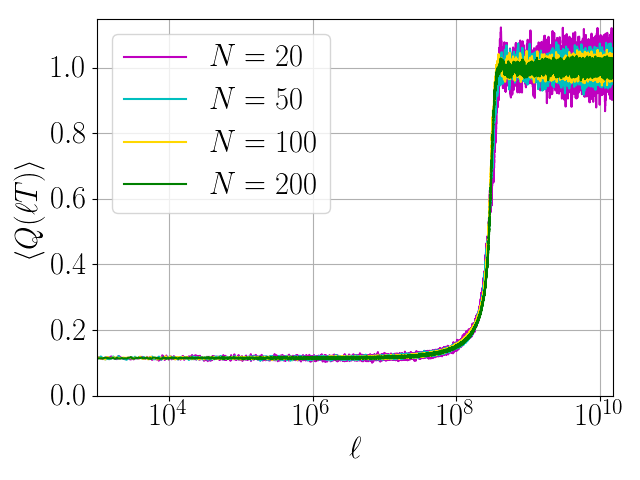}
			
		\end{tabular}

		\caption{Average energy absorption $\langle Q(t) \rangle$ as a function of the number of driving cycles $\ell$ for system sizes $N=20,50,100,200$. The graphs shown have driving frequencies $\Omega/J = 3.8$ (left), $\Omega/J = 4.2$ (middle) and $\Omega/J = 4.6$ (right). The simulation parameters are $h/J = 0.809$, $g/J = 0.9045$. Note how the fluctuations decrease with system size while the steepness increases with system size. }
		\label{fig:size_dep}
		
	\end{figure}

	\section{\label{sec:numerical_error}Stability of the Numerical Integration of the Hamilton Equations}
	
	The Hamilton EOM for our model can be solved using an analytic recursion relation during in each half period. This means that the stroboscopic evolution of the system is described by an iterative application of a discrete nonlinear map [see main text]. Iterative recursive relations are easier to simulate numerically than a system of time-dependent non-linear differential equations. 
	
	Nevertheless, when we deal with times as large as $10^{10}$ driving cycles, it is imperative to ensure that the numerical error growth is under control at all times. To test it, recall that the spin algebra requires that the sum of the squared on-site magnetizations
	\begin{equation}
	C_{j}( \ell T) = \vert\vec S_{j}(\ell T))\vert^{2}
	\end{equation}
	is an integral of motion. In particular, it is fixed to unity for every lattice site and at all times. The origin of this integral of motion can be traced back to the Casimir operator of the rotor algebra.
	
	The relative deviation of $C_{j}(\ell T)$ from unity represents a reliable measure of how large the numerical integration error is. Since the squared magnetization is defined for each lattice site, an absolute value for the error is given by the maximum deviation over the entire chain:
	\begin{equation}
	C(\ell T) = \text{max}_{j} |C_{j}(\ell T) - 1|. 
	\end{equation}
	
	Figure~\ref{fig:norm} shows the time-dependence of $C(\ell T)$. One can see that the maximum error is no bigger than $10^{-5}$ at the end of the time evolution. This is a good indicator that the numerical simulation is sufficiently accurate to render our analysis reliable. Indeed, note that (exponential) blow up of the curves due to numerical error happens at times order of magnitudes larger than stage (iii) [see Fig.~1, main text], during which unconstrained thermalization to an infinite-temperature state occurs for a given frequency. Nevertheless, the pre-factor is so small that this blow-up is under control for all times of interest. 
	
	For comparison, the average error $N^{-1}\sum_{j=1}^N |C_{j}(\ell T) - 1|\sim  10^{-10}$ is a few orders of magnitude smaller at $\ell=10^10$, due to self-averaging effects.
	
	\begin{figure}[t!]
		\centering
		\includegraphics[width=0.5\textwidth]{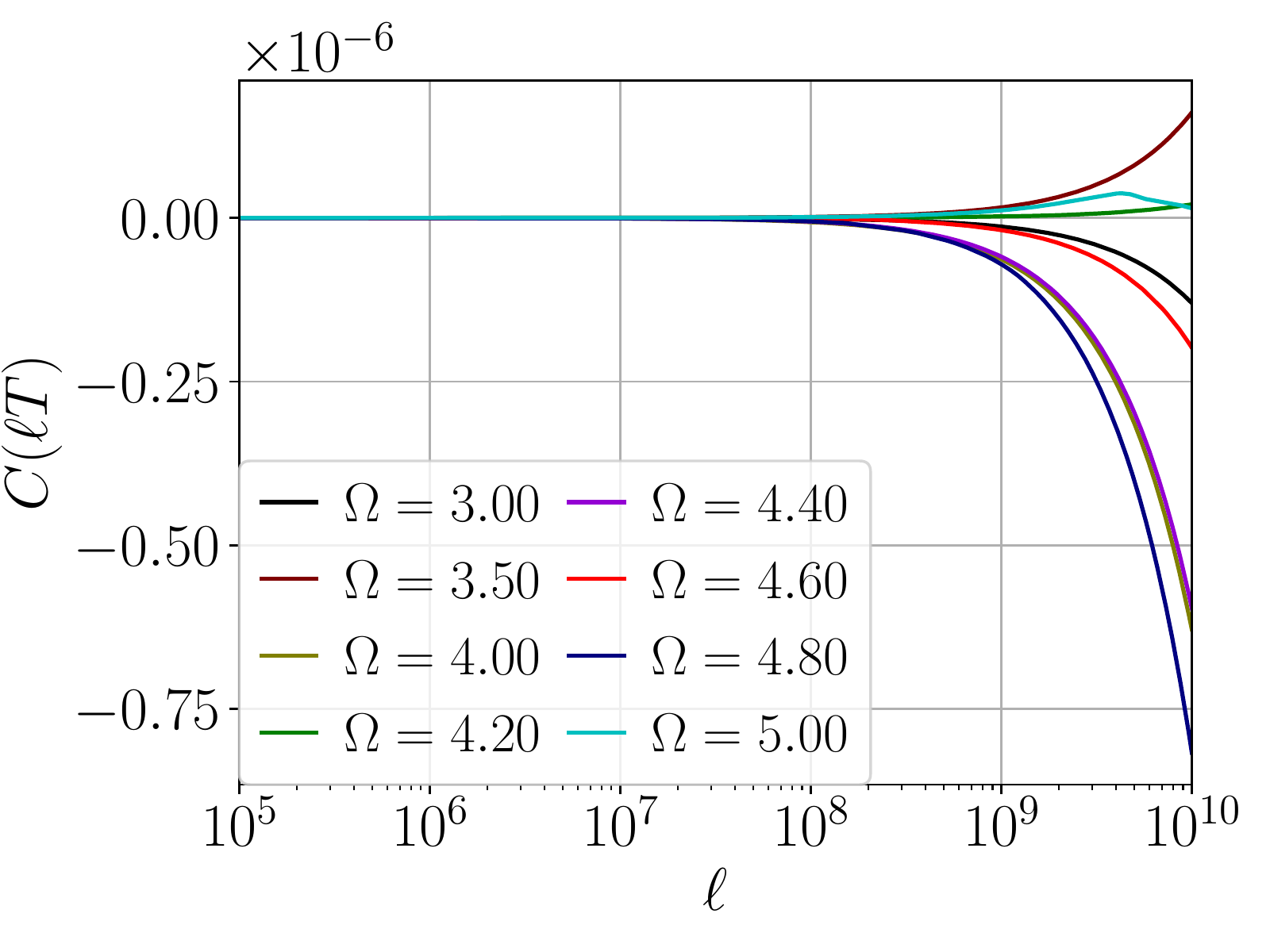}
		\caption{The maximum error of the Casimir operator $C(\ell T)$ as a function of the number of driving cycles $\ell$ for different $\Omega/J$ values.  The simulation parameters are $h/J = 0.809$, $g/J = 0.9045$, and $N = 100$.}
		\label{fig:norm}
	\end{figure}
	
	\section{\label{sec:noise_stabiliry}Dependence on the Size of the Noise Distribution for the Initial State}
	
	As we explained in the main text, in order to make our results robust to the chaotic character of many-body dynamics, we add small noise in the azimuthal angle of the initial antiferromagnetic state. By studying the long time behavior of energy absorption, here we show that the size of the support of the noise distribution is irrelevant to the qualitative picture of the thermalizing dynamics, provided the energy of the initial state remains unaffected. Figure~\ref{fig:noise} shows the time-dependence of the energy absorption for three different values of the support of the uniform noise distribution: $[-\pi/W,\pi/W]$, with $W\in\{50,150,500\}$.
	
	\begin{figure}[t!]
		
		\begin{tabular}{ll}
			\includegraphics[width=0.333\columnwidth]{{noise-strength_Omega=3.8}.png}
			\includegraphics[width=0.333\columnwidth]{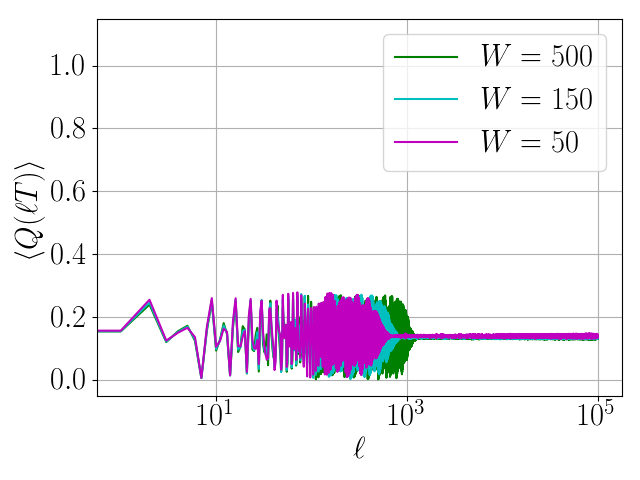}
			\includegraphics[width=0.333\columnwidth]{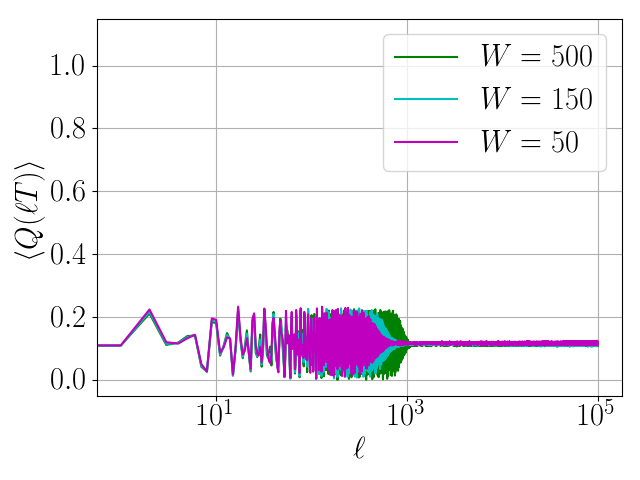}
		\end{tabular}
		
		\begin{tabular}{ll}
			\includegraphics[width=0.333\columnwidth]{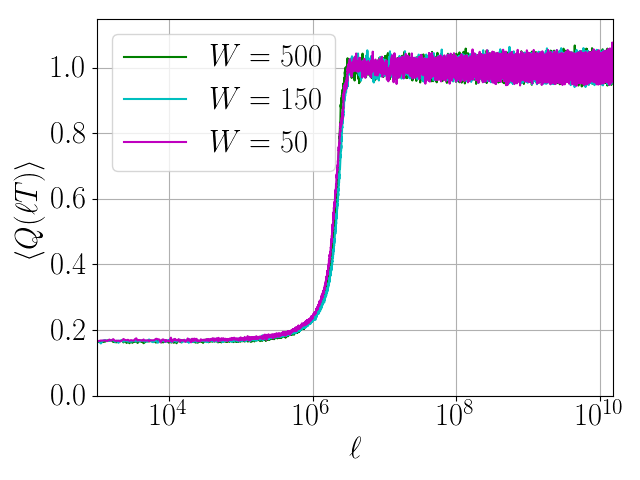}
			\includegraphics[width=0.333\columnwidth]{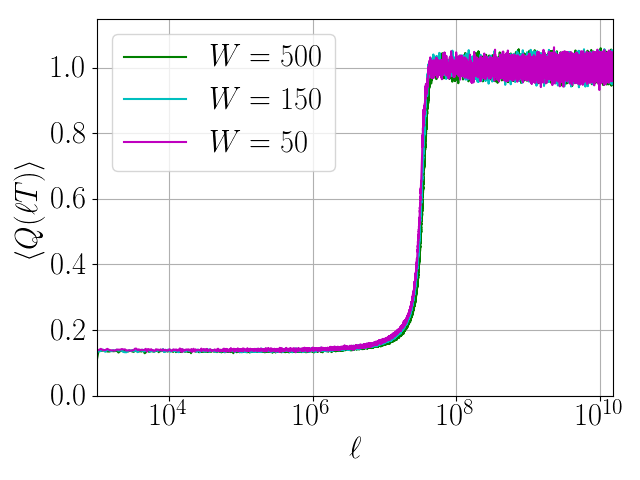}
			\includegraphics[width=0.333\columnwidth]{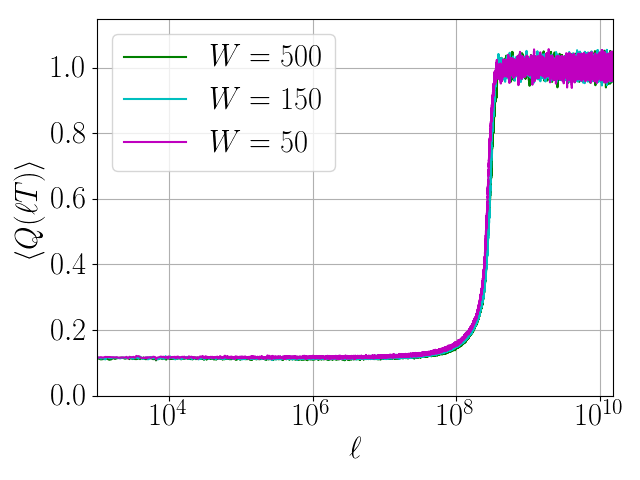}
		\end{tabular}
		
		\centering
		\caption{Average energy absorption $\langle Q(t) \rangle$ as a function of the number of driving cycles $\ell$, for $\Omega/J = 3.8$ (left), $\Omega/J = 4.2$ (middle) and $\Omega/J = 4.6$ (right). The above panels show the initial fluctuations due to two particle behavior and the lower show behavior for long times. The $W$ value corresponds to the strength of the initial noise.The simulation parameters are $h/J = 0.809$, $g/J = 0.9045$, and $N = 100$.}
		\label{fig:noise}
		
	\end{figure}

\end{widetext}

\end{document}